\documentclass[notitlepage,a4paper,aps,prd,onecolumn,nofootinbib,groupedaddress]{revtex4}
\usepackage{amsmath}
\usepackage{amsfonts,color}
\usepackage{amsthm}

\usepackage{pdfpages}
\usepackage{amsfonts,color}
\usepackage{amssymb,float}
\usepackage{cancel}
\usepackage{empheq}
\newcommand{\lc}[1]{\accentset{\circ}{#1}}

\usepackage{amsmath,accents}

\usepackage[utf8]{inputenc}
\allowdisplaybreaks
\usepackage{color}
\usepackage{hyperref}
\hypersetup{
    colorlinks=true,
    linkcolor=blue,
    filecolor=magenta,      
   citecolor=blue
}

\usepackage{enumitem}

\usepackage{tikz}
\usetikzlibrary{shapes.geometric}
\usetikzlibrary{arrows.meta,arrows}

\newcommand{\de}{\delta}
\newcommand{\om}{\omega}
\newcommand{\Tom}{\Tilde{\omega}}
\newcommand{\eps}{\epsilon}
\newcommand{\pd}{\partial}
\newcommand{\Lagr}{\mathcal{L}}

\newcommand{\dd}{{\rm d}}
\newcommand{\DD}{{\rm D}}

\begin{document}
\title{Symmetric Teleparallel Gauss-Bonnet Gravity and its Extensions}

\author{Juan Manuel Armaleo$^{a,b}$}
\email{jarmaleo@df.uba.ar}

\author{Sebastian Bahamonde$^{c,d}$}
\email{sbahamondebeltran@gmail.com, bahamonde.s.aa@m.titech.ac.jp}

\author{Georg Trenkler$^{e,f}$}
\email{trenkler@fzu.cz}

\author{Leonardo G. Trombetta$^{e}$}
\email{trombetta@fzu.cz}

\affiliation{${}^{a}$Universidad de Buenos Aires, Facultad de Ciencias Exactas y Naturales, Departamento de F\'isica, 1428 Buenos Aires, Argentina\\
${}^{b}$CONICET - Universidad de Buenos Aires, Instituto de F\'isica de Buenos Aires (IFIBA). Buenos Aires, Argentina
Ciudad Universitaria, Pabellon I, 1428 Buenos Aires, Argentina.\\
${}^{c}$Department of Physics, Tokyo Institute of Technology 2-12-1 Ookayama, Meguro-ku, Tokyo 152-8551, Japan.\\
${}^{d}$Kavli Institute for the Physics and Mathematics of the Universe (WPI), The University of Tokyo Institutes
for Advanced Study (UTIAS), The University of Tokyo, Kashiwa, Chiba 277-8583, Japan.\\
${}^{e}$CEICO, Institute of Physics of the Czech Academy of Sciences, Na Slovance 1999/2, 182 21, Prague 8, Czechia.\\
${}^{f}$Institute of Theoretical Physics, Faculty of Mathematics and Physics, Charles University, V Holešovičkách 2, 180 00 Prague 8, Czechia.}

\begin{abstract}
General Teleparallel theories assume that curvature is vanishing in which case gravity can be solely represented by torsion and/or nonmetricity. Using differential form language, we express the Riemannian Gauss-Bonnet invariant concisely in terms of two General Teleparallel Gauss-Bonnet invariants, a bulk and a boundary one. Both terms are boundary terms in four dimensions. We also find that the split is not unique and present two possible alternatives. In the absence of nonmetricity our expressions coincide with the well-known Metric Teleparallel Gauss-Bonnet invariants for one of the splits. Next, we focus on the description where only nonmetricity is present and show some examples in different spacetimes. We finish our discussion by formulating novel modified Symmetric Teleparallel theories constructed with our new scalars.\end{abstract}

\maketitle

\section{Introduction}

General Relativity (GR) is the flagship theory of gravity and has been very successful in describing gravitational physics as well as predicting the existence of gravitational waves, black holes and an expanding universe. It is however not without its limitations, as many aspects remain unclear, such as the unavoidable appearance of singularities, the nature of Dark Matter and Dark Energy and the lack of a compelling UV completion. One way to address some of these is to modify the theory by changing some of its underlying assumptions. Lovelock's theorem ensures the Einstein-Hilbert action to uniquely describe (pure) gravity in Riemannian geometry with field equations that are at most second order in $D=4$ dimensions. The next invariant that appears in an expansion in powers of the curvature is the Gauss-Bonnet invariant~\cite{Lovelock:1972vz,Lovelock:1971yv},
\begin{equation}\label{RiemGB}
    \lc{G} = \lc{R}_{\mu\nu\rho\sigma} \lc{R}^{\mu\nu\rho\sigma} - 4 \lc{R}_{\mu\nu} \lc{R}^{\mu\nu} + \lc{R}^2,
\end{equation}
which is a topological invariant in $D=4$, therefore not affecting the dynamics. However, precisely because of its topological character, it can lead to interesting features in a theory of gravity. In order to make this quantity relevant in $D=4$ it is necessary instead to let go some of the other assumptions of Lovelock's theorem. For instance, by allowing for field equations of higher-than-second order nature, which generically imply the appearance of new degrees of freedom (d.o.f.), one can then construct a theory containing the Gauss-Bonnet invariant in a nontrivial way, such as $f(\lc{R},\lc{G})$ gravity~\cite{Nojiri:2005vv,Nojiri:2005jg,Nojiri:2017ncd,Cognola:2006eg,Li:2007jm,Nojiri:2018ouv}. 

Another possibility to modify GR is coupling $\lc{G}$ directly to new degrees of freedom. Indeed, in the context of scalar-tensor theories of gravity in Riemannian geometry, a lot of attention has been given in recent years to a coupling between the Gauss-Bonnet invariant and the scalar field $\psi$ of the form $f(\psi) \lc{G}$, which can be generated for example at low energies by string theory~\cite{Fradkin:1984pq,Gross:1986mw,Callan:1985ia}, or in general by the spontaneous breaking of a conformal symmetry, with $\psi$ related with the corresponding dilaton \cite{Komargodski:2011vj,Luty:2012ww}. This type of operator belongs to the Horndeski class of theories, therefore leading to second-order field equations for both the scalar and the metric \cite{Kobayashi:2011nu}, and also inheriting nice properties such as nonrenormalization \cite{Pirtskhalava:2015nla}. Phenomenological implications in cosmology include for example the existence of eternal solutions with zero cosmological constant \cite{Armaleo:2017lgr} and inflationary models where the inflaton is nonminimally coupled to the Gauss-Bonnet invariant \cite{Carter:2005fu,Pozdeeva:2020apf,Odintsov:2023weg}. On the astrophysical side, the main appeal is the fact that the presence of such operator generically induces scalarization of black-hole solutions \cite{Sotiriou:2013qea,Sotiriou:2014pfa}, which include interesting scenarios of spontaneous scalarization \cite{Silva:2017uqg,Macedo:2019sem} (also spin-induced \cite{Dima:2020yac,Herdeiro:2020wei}) when the coupling is non-linear in $\psi$.\footnote{The coupling leading to spontanous scalarization of black holes has also been shown to trigger instabilities in the context of inflationary cosmology~\cite{Anson:2019uto}.} 

In the special case of a linear coupling $\psi \, \lc{G}$, interestingly the theory enjoys shift-symmetry\footnote{Due to the topological character of $\lc{G}$ in $D=4$, a constant shift $\psi \to \psi + c$ does not affect the dynamics.} and has been shown to be the \emph{only} way of evading known no-hair theorems for shift-symmetric scalar-tensor (Riemannian) theories and flat asymptotics \cite{Hui:2012qt}, even when considering the larger Degenerate Higher-Order Scalar-Tensor (DHOST) class of theories \cite{Creminelli:2020lxn}. Black holes in this case are always scalarized, which implies strong constraints from gravitational-wave observations due to unavoidable scalar-wave emission during the inspiral phase of binary mergers \cite{Yagi:2012gp,Yunes:2016jcc,Witek:2018dmd,Lyu:2022gdr}, although they could still lead to potentially observable effects in the ringdown phase and provide insight about Dark Energy scenarios \cite{Noller:2019chl}. Strong constraints also follow from theoretical consistency considerations such as the requirement of causality \cite{Serra:2022pzl}. 

Yet another way of modifying GR is by changing the underlying geometrical notion used to describe gravity. It has been shown that it is possible to formulate GR equivalently either in terms of curvature, torsion or nonmetricity \cite{Nester:1998mp,BeltranJimenez:2019tjy,Heisenberg:2018vsk}. The scenario in which the general curvature is switched off and still the theory provides the Einstein's field equations is known as the General Teleparallel equivalent of GR theory~\cite{Obukhov:2002tm,BeltranJimenez:2018vdo,BeltranJimenez:2017tkd}. Modifications of that theory are however different depending on which geometrical formulation is chosen. When not only curvature is zero but also nonmetricity(torsion) is vanishing, one restricts the theories to be part of the so-called metric(symmetric) Teleparallel theories or torsional(nonmetricity-based) Teleparallel theories~\cite{Aldrovandi:2013wha,Maluf:2013gaa,Krssak:2015oua,Cai:2015emx,Krssak:2018ywd,Bahamonde:2021gfp,BeltranJimenez:2019tjy}. For simplicity, we will use the abbreviation TG for metric (or torsional) Teleparallel gravity, while Symmetric TG will refer to Symmetric Teleparallel gravity when only nonmetricity is present. Furthermore, the abbreviation General TG will refer to the General Teleparallel theories with vanishing curvature and non-zero torsion and nonmetricity.

In particular, any Teleparallel theory could allow for novel ways of including the Gauss-Bonnet invariant in an action nontrivially. 
Most of the known results in black hole physics or cosmology assume Riemannian geometry and the conclusions no longer apply in this case. It is interesting to ask whether they would still hold when considering instead a Teleparallel formulation. In order to pursue these studies it is necessary to have manageable expressions for the Riemannian Gauss-Bonnet invariant in terms of teleparallel quantities. For the case of TG this has been already developed in Ref.~\cite{Kofinas:2014owa}, where the Riemannian Gauss-Bonnet invariant is split into two terms: a bulk term denoted as $T^{(T)}_G$, and a boundary one as $B^{(T)}_G$. Those terms allow us to construct modified Metric TG theories with non-trivial dynamics for the Metric TG Gauss-Bonnet invariants, see~\cite{Kofinas:2014owa,Bahamonde:2016kba,Gonzalez:2015sha} as examples. There are interesting works related to those torsional theories in the context of cosmology~\cite{Kofinas:2014aka,Kofinas:2014daa,Capozziello:2016eaz,Bahamonde:2018ibz,delaCruzDombriz:2017lvj,Bahamonde:2020vfj}, and recently, in black hole physics~\cite{Bahamonde:2022chq}. It has been found that the Metric TG Gauss-Bonnet invariants provided by torsion generate scalarized black-hole solutions with spontenous scalarization, and in general, they have different features as in the Riemannian case~\cite{Bahamonde:2023llu}.

In this paper we aim to extend the work of Ref.~\cite{Kofinas:2014owa} to the General TG framework, where both torsion and nonmetricity are present, by formulating a compact and manageable expression for the Riemannian Gauss-Bonnet invariant $\lc{G}$. As a special case, this includes the so-called Symmetric TG where only nonmetricity is different from zero.  In this latter setup, which has been so far the least explored case in the literature, we then use our results to discuss some novel formulations of Symmetric TG theories that include the Gauss-Bonnet invariant. 

The paper is organised as follows: In Sec.~\ref{sec:symmetricTG} we introduce General TG geometries using both differential form language and their corresponding tensorial expressions. We also formulate the General TG equivalent of GR. Then, in Sec.~\ref{sec:GB} we split the Riemannian Gauss-Bonnet invariant into two General TG scalars which are both boundary terms in 4-dimensions. Sec.~\ref{sec:STG} is devoted to studying the particular case of Symmetric TG where we build  their Gauss-Bonnet invariants, construct different new theories, and explore some examples for different spacetimes. We summarise and discuss our main results in Sec.~\ref{sec:conclusions}.

The notation throughout this paper is as follows: Latin indices $a, b,\dots$ run over the tangent space of the $D$-dimensional spacetime while Greek indices $\mu,\nu,\dots$ run over all coordinates of the $D$-dimensional spacetime. Quantities denoted with tildes on top are referring to a general connection, whereas quantities with an overcircle denote the Riemannian ones (computed using the Levi-Civita connection). Further, quantities without any symbol on top are of purely Teleparallel nature.

\section{General Teleparallel Geometries}\label{sec:symmetricTG}
In this section we revise the formalism of differential forms and introduce the relevant geometrical quantities. For this purpose, we mainly follow the work presented in \cite{Kofinas:2014owa, Adak:2023ymc, Gasperini:2017ggf} and encourage the reader to see the aforementioned references for more details. We list the definitions of the geometrical quantities in the language of differential forms needed to construct the General TG equivalent of GR.
\subsection{Geometrical quantities in differential form}

In the early 1920's Élie Cartan presented an alternative formulation of GR in terms of two new dynamical quantities: the (dual) \textit{vielbein} $e^a$, also known as\footnote{The formal definition of the \textit{vielbein} is with the lower index $e_a$ while the dual vielbein is defined as $e^a$. The implicit notation is understood depending whether the index is up or low. Both objects are related \textit{via} the dual property $e_a e^b=\de^a_b$ where $\de^a_b$ is the Kronecker delta.} the \textit{basis of the cotangent space} or \textit{tetrads}, and the \textit{connection 1-form} $\Tom^a{}_b$, also known as \textit{spin connection}. As opposed to Einstein's idea, Cartan's idea was to distinguish between the metric and the connection as two different objects (unlike GR in which the Levi-Civita connection is assumed). The connection then is not written in terms of the metric anymore but they are independent. The information about the geometry (\textit{i.e.} gravitation) which in GR is encoded in the Ricci scalar $\tilde{R}=g^{\mu\nu}\tilde{R}_{\mu\nu}$ \textit{via} the metric, will now be given by the three quantities $(Q_{ab},T^a, \tilde{R}^a{}_b)$ known as the \textit{nonmetricity 1-form}, the \textit{torsion 2-form} and the \textit{curvature 2-form} respectively. They are defined as
\begin{eqnarray}
    Q_{ab} &:=&  \tilde{\DD} g_{ab} := \dd g_{ab}-\Tom_{ab}-\Tom_{ba}\,,\\
    T^a &:=& \tilde{\DD} e^a := \dd e^a+\Tom^a{}_c\wedge e^c \,,\label{eq:torsion-1-form} \\
    \Tilde{R}^a{}_b &:=& \dd\Tom^a{}_b+\Tom^a{}_c\wedge\Tom^c{}_b\,,
\end{eqnarray}
where $\dd$ is the exterior derivative and the symbol $\wedge$ denotes the wedge product; $g_{ab}$ stands for the metric tensor, with which we raise and/or lower indices. The above equations are known as \textit{Cartan's structure equations}, and they define the nonmetricity 1-form $Q_{ab}$, the torsion 2-form $T^a$ and the curvature 2-form $\Tilde{R}^a{}_b$ in terms of the basis $e^a$ and the connection 1-form $\Tom^a{}_b$. The \textit{general covariant exterior derivative} $\tilde{\DD}$ is defined in the following way: acting on a set of $p-$forms $\Phi^a{}_b$, 
\begin{equation}
   \tilde{ \DD}\Phi^a{}_b = \dd \Phi^a{}_b + \Tom^a{}_c\wedge\Phi^c{}_b - \Tom^c{}_b\wedge\Phi^a{}_c\,.
\end{equation}
Note that the covariant exterior derivative $\tilde{\DD}$ acts in the same way as the exterior derivative $\dd$ in the sense that when applied on a $p-$form, it gives a $(p+1)-$form. The connection terms ensure that $\tilde{\DD}$ (when applied to a $p-$form) transforms as a tensorial object in the same way as the Levi-Civita connection $\lc{\Gamma}^a{}_{bc}$ ensures that the covariant derivative $\lc{\nabla}$ behaves like a tensor when applied to a tensor field in GR. The connection 1-form $\Tom^a{}_b$ defines the parallel transportation. With the covariant exterior derivative defined, we can write the Bianchi identities as
\begin{align} \label{eq:Bianchi}
  \Tilde{R}^a{}_b\wedge e^b=\tilde{\DD} T^a\,,\qquad \tilde{\DD} \Tilde{R}^a{}_b=0 \,,\qquad   (\Tilde{R}_{ab}+\Tilde{R}_{ba})=-\tilde{\DD} Q_{ab}\,.
\end{align}

The affine connection 1-form can be decomposed into Riemannian (\textit{i.e} Levi-Civita) and non-Riemannian parts according to $\Tom^a{}_b=\lc{\om}^a{}_b+N^a{}_b$. Here $\lc{\om}^a{}_b$ stands for the Levi-Civita (or Christoffel) connection 1-form, satisfying $\lc{\om}_{ab}=-\lc{\om}_{ba}$, and $N^a{}_b$ is known as the \textit{distortion 1-form}. As can be seen, the distortion $N$ quantifies the difference between the Levi-Civita connection $\lc{\om}$ and the general connection $\Tom$. By replacing the decomposition of the affine connection 1-form into the definition of the curvature 2-form we can get the split of the latter into, again, Riemannian and non-Riemannian parts as
\begin{equation} \label{curvature-split}
    \Tilde{R}^{a}{}_b = \lc{R}^{a}{}_b + \lc{\DD} N^{a}{}_b + N^a{}_c \wedge N^{c}{}_b\,,
\end{equation}
where the Riemannian part of the curvature 2-form is defined, in terms of the Levi-Civita affine connection 1-form, as
\begin{equation}
     \lc{R}^a{}_b:=\dd\lc{\om}^a{}_b+\lc{\om}^a{}_c\wedge\lc{\om}^c{}_b
\end{equation}
and the covariant exterior derivative w.r.t. the Levi-Civita connection acting on the distortion 1-form is given by
\begin{equation}\label{CurvRel}
    \lc{\DD}N^a{}_b=\dd N^a{}_b+\lc{\om}^a{}_c\wedge N^c{}_b-\lc{\om}^c{}_b\wedge N^a{}_c\,.
\end{equation}
The distortion 1-form, moreover, can be decomposed as
\begin{equation}
    N^a{}_b = K^a{}_b + L^a{}_b\,.
\end{equation}
The first term $K^a{}_b$ that satisfies $K_{ab}=-K_{ba}$ is known as the \textit{contortion 1-form} while the second term is the \textit{disformation 1-form} $L^a{}_b$, associated to torsion and nonmetricity respectively.

\subsection{Tensorial notation}

When doing computations in differential form language, it will be useful to be able to recover the known expressions written in tensorial form. We will focus on the relevant quantities that are going to appear throughout the work. To be named, we start by stating the tensorial expression for the torsion 2-form,
\begin{equation} \label{eq:torsion-form-to-tensor}
    T^a:=\frac12 T^a{}_{bc}\,e^b\wedge e^c\,,
\end{equation}
In the same way, the expression for the curvature 2-form is given by
\begin{equation} \label{eq:curv-Rmn}
    \tilde{R}^a{}_b:=\frac12 \tilde{R}^a{}_{b\,cd}\, e^c\wedge e^d\,,
\end{equation}
where $\tilde{R}^a{}_{b\,cd}$ are the components of the general Riemann tensor. It will be useful to keep in mind that the exterior derivative, when acted on an object, can be written as $d(\cdot)=\pd_d(\cdot)e^d$. Finally, the distortion 1-form, when written in its tensorial form, reads
\begin{eqnarray} \label{eq:distorsion-form-to-tensor}
    N^a{}_b:=N^a{}_{b\,c}\,e^c\,.
\end{eqnarray}
When moving from the orthonormal to coordinate basis, it is important to remember the ``tetrad postulate"
\begin{equation} \label{tetrad-postulate}
    \tilde{\nabla}_\mu e^a{}_{\nu} = 0, 
\end{equation}
which is valid for \emph{any} connection.

Finally we write down the well known expressions in the coordinate basis for the objects defined above. We start by recalling the expressions for the curvature, torsion and nonmetricity, which define the Cartan's structure equations,
\begin{eqnarray}
   \tilde{R}^{\sigma}\,_{\rho\mu\nu} &=& \partial_{\mu}\tilde{\Gamma}^{\sigma}\,_{\nu\rho}-\partial_{\nu}\tilde{\Gamma}^{\sigma}\,_{\mu\rho}+\tilde{\Gamma}^{\sigma}\,_{\mu\lambda}\tilde{\Gamma}^{\lambda}\,_{\nu\rho}-\tilde{\Gamma}^{\sigma}\,_{\mu\lambda}\tilde{\Gamma}^{\lambda}\,_{\nu\rho}\,,\\
    T^{\sigma}\,_{\mu\nu} &=& \tilde{\Gamma}^{\sigma}\,_{\mu\nu}-\tilde{\Gamma}^{\sigma}\,_{\nu\mu}\,,\\
     Q_{\rho\mu\nu} &=& \tilde{\nabla}_{\rho} g_{\mu\nu} =\partial_{\rho}g_{\mu\nu}-\tilde{\Gamma}^{\lambda}\,_{\rho\mu}g_{\lambda\nu}-\tilde{\Gamma}^{\lambda}\,_{\rho\nu}g_{\mu\lambda}\,, \label{nonmetricity-tensor}
\end{eqnarray}
In the same way, the distortion tensor $N$ can decomposed into the contortion and disformation tensor as
\begin{eqnarray}
    N^{\lambda}\,_{\mu \nu}=K^{\lambda}\,_{\mu \nu}+L^{\lambda}\,_{\mu \nu}\,,\label{distortion}
\end{eqnarray} 
where, expressed in terms of torsion and nonmetricity, they read~\cite{Hehl:1976kj}:
\begin{align}
    K^{\lambda}\,_{\mu \nu}&=\frac{1}{2}\left(T^{\lambda}\,_{\mu \nu}-T_{\mu}\,^{\lambda}\,_{\nu}-T_{\nu}\,^{\lambda}\,_{\mu}\right)\,,\\
    L^{\lambda}\,_{\mu \nu}&=\frac{1}{2}\left(Q^{\lambda}\,_{\mu \nu}-Q_{\mu}\,^{\lambda}\,_{\nu}-Q_{\nu}\,^{\lambda}\,_{\mu}\right)\,. \label{disformation-tensor}
\end{align}
We now have all the ingredients both in differential form and in tensorial language that will allow us to formulate GR in the the Teleparallel framework.

\subsection{General teleparallel formulation of GR}

In GR, i.e. for the Levi-Civita connection, the action is built from the Riemannian Ricci scalar. In the language of differential forms this object is obtained from the curvature 2-form in such a way that the Einstein-Hilbert action (in $D$ dimensions) is written as 
\begin{eqnarray}\label{actionEH}
    \mathcal{S}_{\rm EH} = \frac{1}{2\kappa^2} \int\mathrm{d}^D\!x\, \lc{\Lagr}_{\rm GR}\,,
\end{eqnarray}
where $\kappa^2=8\pi G$ and
\begin{equation}
    \lc{\Lagr}_{\rm GR}= \frac{1}{(D-2)!}\,\epsilon_{a_1...a_D} \,
    \lc{R}^{a_1 a_2} \wedge e^{a_3} \wedge \dots \wedge e^{a_D} = \lc{R} \, *\!1\,,
\end{equation}
is the GR Lagrangian, with $*$ standing for the \textit{Hodge dual operator} that defines the orientation of the manifold
\begin{equation} \label{volume-form}
    *1=\frac{1}{D}\eps_{a_1...a_D}\,e^{a_1}\wedge e^{a_2} \wedge \dots \wedge e^{a_D}\,.
\end{equation}
Here, $\eps_{a_1...a_D}$ is the totally antisymmetric Levi-Civita
symbol (with $\eps_{12...D}=+1$). It is important to note that in Einstein's GR, both torsion and nonmetricity are zero, $T^a=Q^a{}_b=0$, but the curvature it is not, $\lc{R}^a{}_b\neq 0$. On the other hand, one can play with different combinations of the three quantities to build different classes of spacetimes. There are three of such a kind that are going to be of interest for this work: \textit{General Teleparallel geometries} where $Q_{ab}\neq 0, T^a\neq 0, \Tilde{R}_{ab}=0$; \textit{Teleparallel geometries} (torsional) where $T_{ab}= 0, T^a\neq 0, \Tilde{R}_{ab}=0$ and; \textit{Symmetric Teleparallel geometries} where $Q_{ab}\neq 0, T^a=0, \Tilde{R}_{ab}=0$. Teleparallel geometries are characterise by satisfying the so-called \textit{teleparallel condition}:
\begin{eqnarray}\label{eq:tel-cond}
   \Tilde{R}^a{}_b=0\,.
\end{eqnarray}
Note that from this expression, and using \eqref{curvature-split}, we can obtain a relationship between $\lc{R}^a{}_b$ and $N^a{}_b$, which will be useful later on in this work. Also, for teleparallel quantities, we drop any symbol on top.

The above results will be relevant when computing the Gauss-Bonnet invariant in differential forms language, defined in the following section. If, instead, we want to recover the tensorial result as in GR, it will be necessary to keep in mind the expression for the Ricci scalar in the General Teleparalel case \cite{BeltranJimenez:2019odq}:
\begin{eqnarray}\label{RGeneral}
 \tilde{R}&=& \lc{R}+\frac{1}{4}T_{\lambda \mu \nu}T^{\lambda \mu \nu}+\frac{1}{2}T_{\lambda \mu \nu}T^{\mu \lambda \nu}-T^{\lambda}\,_{\lambda\nu}T^{\mu}\,_{\mu}\,^{\nu}+T_{\lambda\mu\nu}Q^{\nu\lambda\mu}+\frac{1}{4}Q_{\lambda\mu\nu}Q^{\lambda\mu\nu}-\frac{1}{2}Q_{\lambda\mu\nu}Q^{\mu\lambda\nu}+\frac{1}{2}Q^{\nu\lambda}\,_{\lambda}Q^{\mu}\,_{\mu\nu}\nonumber\\
    &&-\frac{1}{4}Q^{\nu\lambda}\,_{\lambda}Q_{\nu}\,^{\mu}\,_{\mu}-T^{\lambda}\,_{\lambda\nu}Q^{\nu\mu}\,_{\mu}+\,T^{\lambda}\,_{\lambda\nu}Q^{\mu\nu}\,_{\mu}-2\lc{\nabla}_{\mu}T^{\nu \mu}\,_{\nu}+\lc{\nabla}_{\mu}Q^{\mu}\,_{\nu}\,^{\nu}-\lc{\nabla}_{\mu}Q^{\nu}\,_{\nu}\,^{\mu}\,,\label{BB9}
\end{eqnarray}
from where we can build an action for the General TG equivalent of GR (zero curvature) as~\cite{Bahamonde:2021gfp, BeltranJimenez:2019odq, Iosifidis:2019dua}:
\begin{equation}\label{GTEGR_action}
    \mathcal{S}_{\rm General\, TG-GR} = -\frac{1}{2\kappa^2} \int\mathrm{d}^D\!x \sqrt{-g}\, \mathbb{G}\,,
\end{equation}
with
\begin{eqnarray}
\mathbb{G}&=&  \frac{1}{4}T_{\lambda \mu \nu}T^{\lambda \mu \nu}+\frac{1}{2}T_{\lambda \mu \nu}T^{\mu \lambda \nu}-T^{\lambda}\,_{\lambda\nu}T^{\mu}\,_{\mu}\,^{\nu}+T_{\lambda\mu\nu}Q^{\nu\lambda\mu}+\frac{1}{4}Q_{\lambda\mu\nu}Q^{\lambda\mu\nu}-\frac{1}{2}Q_{\lambda\mu\nu}Q^{\mu\lambda\nu}+\frac{1}{2}Q^{\nu\lambda}\,_{\lambda}Q^{\mu}\,_{\mu\nu}\nonumber\\
    &&-\frac{1}{4}Q^{\nu\lambda}\,_{\lambda}Q_{\nu}\,^{\mu}\,_{\mu}-T^{\lambda}\,_{\lambda\nu}Q^{\nu\mu}\,_{\mu}+\,T^{\lambda}\,_{\lambda\nu}Q^{\mu\nu}\,_{\mu}\,,
\end{eqnarray}
which should not be confused with the Gauss-Bonnet invariant $\lc{G}$. In such a way the Einstein-Hilbert action~\eqref{actionEH} expressed in terms of the Riemannian Ricci scalar 
\begin{equation}
     \mathcal{S}_{\rm EH} = \frac{1}{2\kappa^2} \int\mathrm{d}^D\!x \sqrt{-g}\, \lc{R}
\end{equation}
differs only by a boundary term compared to Eq.~\eqref{GTEGR_action}, meaning that Einstein's field equations arise from both actions. For this reason the theory is labelled as equivalent to GR. One can then enforce conditions on the geometry such that one restricts to the torsional or nonmetricity case. If, for example, we set the torsion to zero, we will get the action for the Symmetric TG case, which will be the dedicated study in Sec.~\ref{sec:STG}.


\section{Gauss-Bonnet invariant in a General Teleparallel geometry}\label{sec:GB}

We now turn our attention to the computation of the Riemannian Gauss-Bonnet invariant in terms of the distortion tensor $N_{\lambda\mu\nu}$ in the General TG case. This object is defined as Eq.~\eqref{RiemGB} which in differential-form language in $D$ dimensions can be obtained as
\begin{equation} \label{RiemGB-form}
    \lc{G} \, *\!1 = \frac{1}{(D-4)!} \epsilon_{a_1...a_D} \lc{R}^{a_1 a_2} \wedge \lc{R}^{a_3 a_4} \wedge e^{a_5} \wedge \dots \wedge e^{a_D},
\end{equation}
in terms of the Riemannian curvature 2-form $\lc{R}^{a}{}_b$, which in turn can be expressed in terms of the distortion tensor $N_{\lambda\mu\nu}$ under the teleparallel condition Eq.~\eqref{eq:tel-cond}. This would be a rather straight-forward enterprise, however we are also interested in identifying within the resulting expression bulk and boundary contributions. As we will see, this split is not unique.

\subsection{Derivation in differential form language} \label{sec:derivation}

We begin first with an approach to the computation that follows closely Ref.~\cite{Kofinas:2014owa}, were the Metric TG equivalent of the Gauss-Bonnet invariant $\lc{G}$ was first derived. The main difference here is that we will let nonmetricity to also be present, in addition to torsion. For this we must work with the distortion 1-form $N^{ab}$ which has a priori no symmetry properties under the exchange of $a$ and $b$, unlike the purely torsional case where the contortion 1-form $K^{ab}$ is anti-symmetric. This requires some care when going through the derivation, and expectedly, some new contributions will arise from the purely symmetric parts of the distortion 1-form, which are uniquely associated with nonmetricity. Notice that nonmetricity also contributes to the anti-symmetric part of $N^{ab}$.

Let us start from the Lagrangian containing the Gauss-Bonnet invariant constructed with the general curvature
\begin{equation} \label{L2}
    \tilde{\Lagr}_{2}= \frac{1}{(D-4)!}\epsilon_{a_1...a_D} \tilde{R}^{a_1 a_2} \wedge \tilde{R}^{a_3 a_4} \wedge e^{a_5} \wedge \dots \wedge e^{a_D} = \tilde{G} \, *\!1\,,
\end{equation}
and replace both factors of the general curvature 2-form $\tilde{R}^{a}{}_{b}$ by the Riemmanian curvature form $\lc{R}^{a}{}_{b}$ and the distortion 1-form $N^a{}_b$ using \eqref{curvature-split}. One obtains
\begin{equation} \label{L2split1}
    (D-4)!\tilde{\Lagr}_{2}= (D-4)!\lc{\Lagr}_{2}+I_{1}+2I_{2}+2I_{3}+2I_{4}+I_{5}\,,
\end{equation}
where
\begin{eqnarray}
    I_{1}&=&\epsilon_{a_1...a_D} N^{a_1}{}_c\wedge N^{c a_2}\wedge N^{a_3}{}_d\wedge N^{d a_4}\wedge e^{a_5} \wedge \dots \wedge e^{a_D}\notag \,, \\
    I_{2}&=&\epsilon_{a_1...a_D} \lc{R}^{a_1 a_2}\wedge N^{a_3}{}_d\wedge N^{d a_4}\wedge e^{a_5} \wedge \dots \wedge e^{a_D} \notag\,, \\ 
    I_{3}&=&\epsilon_{a_1...a_D} \lc{\DD}N^{a_1 a_2}\wedge N^{a_3}{}_d\wedge N^{d a_4}\wedge e^{a_5} \wedge \dots \wedge e^{a_D} \notag\,, \\
    I_{4}&=&\epsilon_{a_1...a_D} \lc{\DD}N^{a_1 a_2}\wedge \lc{R}^{a_3 a_4}\wedge e^{a_5} \wedge \dots \wedge e^{a_D} \notag \,,\\
    I_{5}&=&\epsilon_{a_1...a_D} \lc{\DD}N^{a_1 a_2}\wedge \lc{\DD}N^{a_3 a_4}\wedge e^{a_5} \wedge \dots \wedge e^{a_D}\,,
\end{eqnarray}
and
\begin{equation} \label{RiemL2}
    \lc{\Lagr}_{2}= \frac{1}{(D-4)!}\epsilon_{a_1...a_D} \lc{R}^{a_1 a_2} \wedge \lc{R}^{a_3 a_4} \wedge e^{a_5} \wedge \dots \wedge e^{a_D} = \lc{G} \, *\!1\,,
\end{equation}
which is what we ultimatelly want to compute. Using properties \eqref{lc-D-e}, \eqref{lc-Bianchi-R} and \eqref{lc-D-epsilon}, we readily identify the contributions to what will be the boundary term. In particular, we can rewrite
\begin{eqnarray}\label{I4rew}
    I_{4}&=& \dd\left(\epsilon_{a_1...a_D}N^{a_1 a_2}\wedge \lc{R}^{a_3 a_4}\wedge e^{a_5} \wedge \dots \wedge e^{a_D}\right)\,,
\end{eqnarray}
and
\begin{eqnarray}\label{I5rew}
    I_{5}&=& \lc{\DD}\left(\epsilon_{a_1...a_D}N^{a_1 a_2}\wedge \lc{\DD}N^{a_3 a_4}\wedge e^{a_5} \wedge \dots \wedge e^{a_D}\right)+\epsilon_{a_1...a_D}N^{a_1 a_2}\wedge \lc{\DD}^{2}N^{a_3 a_4}\wedge e^{a_5} \wedge \dots \wedge e^{a_D}    \notag \\ 
    &=& \dd\left(\epsilon_{a_1...a_D}N^{a_1 a_2}\wedge \lc{\DD}N^{a_3 a_4}\wedge e^{a_5} \wedge \dots \wedge e^{a_D}\right)+2\epsilon_{a_1...a_D}N^{a_1 a_2}\wedge \lc{R}^{a_3}{}_c\wedge N^{[c a_4]} \wedge e^{a_5} \wedge \dots \wedge e^{a_D} \,,
\end{eqnarray}
where we integrated by parts in the first line and used property \eqref{D2-N-up-down} in the second line of Eq. \eqref{I5rew}. Hence, the Lagrangian in Eq. \eqref{L2split1} can be further decomposed as
\begin{equation} \label{L2split2}
    (D-4)!\tilde{\Lagr}_{2}= (D-4)!\lc{\Lagr}_{2}+I_{1}+2(I_{3}+I_{6})+\dd B \,,
\end{equation}
where
\begin{eqnarray}
    I_{6}&=&\epsilon_{a_1...a_D} \left(\lc{R}^{a_1 a_2}\wedge N^{a_3}{}_d\wedge N^{d a_4}+N^{a_1 a_2}\wedge \lc{R}^{a_3}{}_c\wedge N^{[c a_4]} \right)\wedge e^{a_5} \wedge \dots \wedge e^{a_D}\notag \,,\\
    B&=&\epsilon_{a_1...a_D}\left(2N^{a_1 a_2}\wedge \lc{R}^{a_3 a_4}  +N^{a_1 a_2}\wedge \lc{\DD}N^{a_3 a_4} \right)e^{a_5} \wedge \dots \wedge e^{a_D}\,.
\end{eqnarray}
Therefore, $I_4$ and one part of $I_5$ gave rise to the boundary term $B$, while $I_2$ and the second part of $I_5$  have been combined into $I_6$. Next, we realize that
\begin{eqnarray}\label{I3plusI6}
    2(I_3 + I_{6})&=&2\epsilon_{a_1...a_D}\biggl[ \left(\lc{\DD}N^{a_1 a_2}+\lc{R}^{a_1 a_2}\right)\wedge N^{a_3}{}_d\wedge N^{d a_4}+N^{a_1 a_2}\wedge \lc{R}^{a_3}{}_c\wedge N^{[c a_4]}\biggr]\wedge e^{a_5} \wedge \dots \wedge e^{a_D} \notag \\
    &=&2\epsilon_{a_1...a_D}\left(\tilde{R}^{a_1 a_2}\wedge N^{a_3}{}_d\wedge N^{d a_4}  - N^{a_1}{}_c\wedge N^{c a_2}\wedge N^{a_3}{}_d \wedge N^{d a_4}+N^{a_1 a_2}\wedge \lc{R}^{a_3}{}_c\wedge N^{[c a_4]}\right)\wedge e^{a_5} \wedge \dots \wedge e^{a_D} \notag \\
    &=&2J_0-2I_1+2J_1\,,
\end{eqnarray}
where we used \eqref{curvature-split2} in the second line of Eq. \eqref{I3plusI6} and defined 
\begin{eqnarray}
    J_{0}&=&\epsilon_{a_1...a_D} \tilde{R}^{a_1 a_2}\wedge N^{a_3}{}_d\wedge N^{d a_4}\wedge e^{a_5} \wedge \dots \wedge e^{a_D}\,,\notag \\
    J_{1}&=&\epsilon_{a_1...a_D} N^{a_1 a_2}\wedge \lc{R}^{a_3}{}_c\wedge N^{[c a_4]}\wedge e^{a_5} \wedge \dots \wedge e^{a_D}\,,
\end{eqnarray}
such that the Lagrangian in Eq. \eqref{L2split2} reads
\begin{equation} \label{L2split3}
    (D-4)!\tilde{\Lagr}_{2}= (D-4)!\lc{\Lagr}_{2}-I_{1}+2(J_{0}+J_{1})+\dd B\,.
\end{equation}
Finally, we use Eq. \eqref{curvature-split3} to replace the remaining factor of the Riemmanian curvature in $J_1$ and obtain 
\begin{eqnarray}
    J_{1}&=&\epsilon_{a_1...a_D} N^{a_1 a_2}\wedge\left(\tilde{R}^{a_3}{}_c + N^{a_3}{}_d\wedge N^{d}{}_{c} - \DD N^{a_3}{}_{c} \right)\wedge N^{[c a_4]}\wedge e^{a_5} \wedge \dots \wedge e^{a_D}\notag \\
    &=&\hat{J}_0+J_2-J_3\,,
\end{eqnarray}
where we defined 
\begin{eqnarray}
    \hat{J}_{0}&=&\epsilon_{a_1...a_D} N^{a_1 a_2}\wedge \tilde{R}^{a_3}{}_c \wedge N^{[c a_4]}\wedge e^{a_5} \wedge \dots \wedge e^{a_D}\notag \,,\\
    J_{2}&=&\epsilon_{a_1...a_D} N^{a_1 a_2}\wedge N^{a_3}{}_d\wedge N^{d}{}_{c}\wedge N^{[c a_4]}\wedge e^{a_5} \wedge \dots \wedge e^{a_D} \notag \,,\\
    J_{3}&=&\epsilon_{a_1...a_D} N^{a_1 a_2}\wedge \DD N^{a_3}{}_{c}\wedge N^{[c a_4]}\wedge e^{a_5} \wedge \dots \wedge e^{a_D}\,,
\end{eqnarray}
such that we arrive at
\begin{equation} \label{L2split4}
    (D-4)!\tilde{\Lagr}_{2}= (D-4)!\lc{\Lagr}_{2}-I_{1}+2(J_{0}+\hat{J}_{0})+2J_{2}-2J_{3}+\dd B\,.
\end{equation}
Enforcing the teleparallel condition $\tilde{R}^{ab}=0\equiv R^{ab}$, leads to $\tilde{\Lagr}_2 = 0$ on the l.h.s. and $J_0=0=\hat{J}_{0}$ on the r.h.s. of Eq. \eqref{L2split4}, yielding
\begin{equation} \label{L2split5}
    (D-4)!\lc{\Lagr}_{2}= I_{1}-2J_{2}+2J_{3}-\dd B\,,
\end{equation}
or, explicitly
\begin{eqnarray} \label{GB-tele}
    \lc{G} \, *\!1 &=& \frac{1}{(D-4)!} \epsilon_{a_1...a_D} \Biggl[ - \dd \left( 2 N^{a_1 a_2} \wedge \lc{R}^{a_3 a_4} \wedge e^{a_5} \wedge \dots \wedge e^{a_D} + N^{a_1 a_2} \wedge \lc{\DD} N^{a_3 a_4} \wedge e^{a_5} \wedge \dots \wedge e^{a_D} \right) \notag \\    
    && \qquad\qquad\qquad\qquad + \Bigl( 2 N^{a_1 a_2} \wedge \DD N^{a_3}{}_{c} \wedge N^{[c a_4]} - 2 N^{a_1 a_2} \wedge N^{a_3}{}_{d} \wedge N^{d}{}_{c} \wedge N^{[c a_4]} \notag \\
    && \qquad\qquad\qquad\qquad\qquad + N^{a_1}{}_f \wedge N^{f a_2} \wedge N^{a_3}{}_h \wedge N^{h a_4} \Bigr) \wedge e^{a_5} \wedge \dots \wedge e^{a_D} \Biggr]\,.
\end{eqnarray}
This is the main result of this section, expressed in differential form language. In the next subsection we provide an alternative derivation, while in Sec.~\ref{sec:tensorial-GB} we translate the result back to tensorial notation, given in Eq.~\eqref{GB-tele-tensorial}. 

\subsection{Alternative split into bulk and boundary terms} \label{sec:derivation-alt}

We can take a different approach and impose the teleparallel condition~\eqref{eq:tel-cond} from the beginning. It is instructive to do so because it will show that the split between bulk and boundary terms is not unique, even in $D \neq 4$ (in $D=4$ the Gauss-Bonnet invariant is itself a total derivative). 

We start by replacing Eq.~\eqref{curvature-split} with $\tilde{R}^a{}_b=R^{a}{}_{b} = 0$ in one of the factors of $\lc{R}^{ab}$ in Eq.~\eqref{RiemGB-form},
\begin{equation} \label{GB-tele1}
    \lc{G} \, *\!1 = \frac{1}{(D-4)!} \epsilon_{a_1...a_D} \left[ - \lc{\DD} N^{a_1 a_2} \wedge \lc{R}^{a_3 a_4} - N^{a_1}{}_f \wedge N^{f a_2} \wedge \lc{R}^{a_3 a_4} \right] \wedge e^{a_5} \wedge \dots \wedge e^{a_D}\,.
\end{equation}
Here, again by using properties \eqref{lc-D-e}, \eqref{lc-Bianchi-R} and \eqref{lc-D-epsilon}, we see that the first term in Eq.~\eqref{GB-tele1} is a total derivative
\begin{eqnarray}
\epsilon_{a_1...a_D} \lc{\DD} N^{a_1 a_2} \wedge \lc{R}^{a_3 a_4} \wedge e^{a_5} \wedge \dots \wedge e^{a_D} &=& \lc{\DD} \left[ \epsilon_{a_1...a_D} N^{a_1 a_2} \wedge \lc{R}^{a_3 a_4} \wedge e^{a_5} \wedge \dots \wedge e^{a_D} \right] \notag \\
&=& \dd \left[ \epsilon_{a_1...a_D} N^{a_1 a_2} \wedge \lc{R}^{a_3 a_4} \wedge e^{a_5} \wedge \dots \wedge e^{a_D} \right] \notag \\
&=& \epsilon_{a_1...a_D} \dd \left[  N^{a_1 a_2} \wedge \lc{R}^{a_3 a_4} \wedge e^{a_5} \wedge \dots \wedge e^{a_D} \right]\,,
\end{eqnarray}
and therefore we have
\begin{eqnarray} \label{GB-tele2}
    \lc{G} \, *\!1 &=& \frac{1}{(D-4)!} \epsilon_{a_1...a_D} \Biggl[ - \dd \left( N^{a_1 a_2} \wedge \lc{R}^{a_3 a_4} \wedge e^{a_5} \wedge \dots \wedge e^{a_D} \right) - N^{a_1}{}_f \wedge N^{f a_2} \wedge \lc{R}^{a_3 a_4} \wedge e^{a_5} \wedge \dots \wedge e^{a_D} \Biggr]\,.
\end{eqnarray}
We now replace $\lc{R}^{a_3 a_4}$ once more time in the second term,
\begin{eqnarray} \label{GB-tele-alt}
    \lc{G} \, *\!1 &=& \frac{1}{(D-4)!} \epsilon_{a_1...a_D} \Biggl[ - \dd \left( N^{a_1 a_2} \wedge \lc{R}^{a_3 a_4} \wedge e^{a_5} \wedge \dots \wedge e^{a_D} \right) \notag \\    
    && \qquad\qquad\qquad\qquad + \Bigl( N^{a_1}{}_f \wedge N^{f a_2} \wedge \lc{\DD} N^{a_3 a_4} + N^{a_1}{}_f \wedge N^{f a_2} \wedge N^{a_3}{}_h \wedge N^{h a_4} \Bigr) \wedge e^{a_5} \wedge \dots \wedge e^{a_D} \Biggr]\,.
\end{eqnarray}
This very short and simple procedure leads to a different way to split the Gauss-Bonnet invariant into a bulk term and a boundary term.  In Appendix~\ref{app:equiv} we show that this expression is equivalent to Eq.~\eqref{GB-tele}, while the corresponding tensorial expressions can be found in Eq.~\eqref{GB-tele-tensorial-alt}.

In general, the presence of mixed terms schematically of the form $N\wedge N\wedge \lc{\DD}N$, with at least one derivative acting on a specific factor, implies there is no absolute split into bulk and boundary pieces, as one can always trade the position of the derivative at the cost of generating a new boundary contribution. In $D=4$ dimensions the Gauss-Bonnet invariant becomes a topological term, and in particular, a purely boundary term in the form of the divergence of a non-tensorial quantity \cite{Yale:2011usf}. 

\subsection{Tensorial expressions} \label{sec:tensorial-GB}

We are now interested in reading off the scalar $\lc{G}$ from our results of Eqs.~\eqref{GB-tele} and \eqref{GB-tele-alt}, which are $D$-forms. In order to do so, we express both the contortion 1-form explicitly as in Eq.~\eqref{eq:distorsion-form-to-tensor}, as well as the curvature 2-form in terms of the Riemann tensor as given in Eq.~\eqref{eq:curv-Rmn}. From the resulting explicit expression we are able to collect as an overall factor proportional to the volume element $D$-form, Eq.~\eqref{volume-form},
\begin{equation}
    e^{a} \wedge e^{b} \wedge e^{c} \wedge e^{d} \wedge e^{a_5} \wedge \dots \wedge e^{a_D} = \epsilon^{a b c d a_5...a_D} \, *\!1\,,
\end{equation}
and extract the desired scalar expression. For our purposes we will also need to use
\begin{equation}
    \frac{1}{(D-4)!} \epsilon_{a_1...a_D} \epsilon^{a b c d a_5...a_D} = \delta^{a\,b\,c\,d}_{a_1 a_2 a_3 a_4}\,.
\end{equation}

Special care needs to be taken when a term involves the covariant exterior derivative w.r.t. the teleparallel connection; more specifically, $\DD N^{a b}$ which appears in Eq.~\eqref{GB-tele}. When we write the distortion 1-form in terms of the corresponding tensor, Eq.~\eqref{eq:distorsion-form-to-tensor}, we need to differentiate also the basis $e^c$. Thus - from its definition in Eq.~\eqref{eq:torsion-1-form} - a term involving the torsion will emerge:
\begin{align}
    \DD N^a{}_b&=\left(\nabla_d N^a{}_{bc}+\frac12N^a{}_{be}T^e{}_{dc}\right)e^d\wedge e^c\,,
\end{align}
where we also used Eq.~\eqref{eq:torsion-form-to-tensor}. This can again be expressed in terms of the anti-symmetric part of the distortion tensor w.r.t. the last two indices $T^{a}{}_{bc} = - 2 N^{a}{}_{[bc]}$, where we are using the same convention as Ref.~\cite{Kofinas:2014owa}. This is not to be mistaken with other anti-symmetrizations we have used so far involving the first two indices instead. Then, accounting also for the raising of an index under the covariant exterior derivative associated to the teleparallel connection $\DD$, Eq. \eqref{D-N-up-up-to-up-down}, we obtain
\begin{align}
    \DD N^{a b} &= - \left(\nabla_d N^{a b}{}_{c} + N^{a b}{}_{h} N^h{}_{cd} \right) e^c \wedge e^d\,,
\end{align}
where we have dropped the anti-symmetrization square brackets since $c$ and $d$ contract with the anti-symmetric $e^c \wedge e^d$.

With this we can immediately obtain the bulk term which we call ${}^1 T^{(T,Q)}_G$ from Eq.~\eqref{GB-tele}, 
\begin{eqnarray}
    {}^1 T^{(T,Q)}_G &=& \delta^{\mu\,\nu\,\rho\,\sigma}_{\mu_1 \mu_2 \mu_3 \mu_4} \left[ N^{\mu_1}{}_{\alpha\mu}N^{\alpha \mu_2}{}_{\nu}N^{\mu_3}{}_{\beta\rho}N^{\beta \mu_4}{}_{\sigma} -2 N^{\mu_1 \mu_2}{}_{\mu}N^{\mu_3}{}_{\alpha\nu}N^{\alpha}{}_{\beta\rho}N^{[\beta \mu_4]}{}_{\sigma} \right. \notag \\
     && \qquad \qquad \left. + 2 g_{\alpha \beta} N^{\mu_1 \mu_2}{}_{\mu} N^{[\mu_3 \alpha]}{}_{\nu} N^{\beta \mu_4}{}_\gamma N^\gamma{}_{\rho\sigma} + 2 g_{\alpha \beta} N^{\mu_1 \mu_2}{}_{\mu} N^{[\mu_3 \alpha]}{}_{\nu} \nabla_\sigma N^{\beta \mu_4}{}_{\rho} \right. \notag \\
     && \qquad \qquad \left. + 4 g_{\alpha \beta} g_{\gamma \delta} N^{\mu_1 \mu_2}{}_{\mu} N^{[\mu_3 \alpha]}{}_\nu N^{\mu_4 \gamma}{}_{\rho} N^{(\delta \beta)}{}_\sigma \right]\,, \label{TGGeneral1} 
\end{eqnarray}
where we have also switched from the orthonormal basis back to the coordinate basis, for which it is useful to remember the ``tetrad postulate", Eq.~\eqref{tetrad-postulate}. 

Similar manipulations can be applied to the corresponding boundary term of Eq.~\eqref{GB-tele}. We first apply Eq. \eqref{curvature-split2} once in order to eliminate the explicit appeareance of $\DD N^{ab}$, and then by also making use of the properties \eqref{lc-D-epsilon} and \eqref{lc-D-e} one arrives at
\begin{eqnarray}
   {}^{1}\!B^{(T,Q)}_G &= &   \frac{1}{\sqrt{-g}} \, \partial_\mu \! \left[ \sqrt{-g} \, \delta^{\mu \, \nu \, \rho \, \sigma}_{\mu_1 \mu_2 \mu_3 \mu_4} N^{\mu_1 \mu_2}{}_{\nu} \left( N^{\mu_3}{}_{ \lambda \rho} N^{\lambda \mu_4}{}_{\sigma} - \frac{1}{2} \lc{R}^{\mu_3 \mu_4}{}_{\rho \sigma} \right) \right]\, . \label{BGGeneral1}
\end{eqnarray}
The Gauss-Bonnet invariant is the sum of Eqs. \eqref{TGGeneral1} and \eqref{BGGeneral1},
\begin{equation} \label{GB-tele-tensorial}
    \lc{G} = {}^1 T^{(T,Q)}_G + {}^{1}\!B^{(T,Q)}_G\,.
\end{equation}

As discussed previously, the split into bulk and boundary terms is not unique, and in fact we provided an alternative one in the previous subsection, given in Eq.~\eqref{GB-tele-alt}. The corresponding tensorial expressions are
\begin{equation} \label{GB-tele-tensorial-alt}
\lc{G} = {}^{2}T^{(T,Q)}_G+{}^{2}\!B^{(T,Q)}_G\,,
\end{equation}
with
\begin{eqnarray}
  {}^{2}T^{(T,Q)}_G   &= &\delta^{\mu\,\nu\,\rho\,\sigma}_{\mu_1 \mu_2 \mu_3 \mu_4} \left[ N^{\mu_1}{}_{\alpha\mu}N^{\alpha \mu_2}{}_{\nu}N^{\mu_3}{}_{\beta\rho}N^{\beta \mu_4}{}_{\sigma} - N^{\mu_1}{}_{\alpha\mu}N^{\alpha \mu_2}{}_{\nu} \lc{\nabla}_\sigma N^{\mu_3 \mu_4}{}_{\rho} \right]\,,\label{TGGeneral2} \\  
 {}^{2}\!B^{(T,Q)}_G&=&-  \frac{1}{2} \frac{1}{\sqrt{-g}} \, \partial_\mu \! \left[ \sqrt{-g} \, \delta^{\mu \, \nu \, \rho \, \sigma}_{\mu_1 \mu_2 \mu_3 \mu_4} N^{\mu_1 \mu_2}{}_{\nu} \lc{R}^{\mu_3 \mu_4}{}_{\rho \sigma} \right]\,.   
\label{BGGeneral2}
\end{eqnarray}
Theories constructed by breaking down the Gauss-Bonnet invariant and coupling $T^{(T,Q)}_G$ and $B^{(T,Q)}_G$ independently will be sensitive to the choice of splitting and will in general not be equivalent. They may coincide accidentally when evaluated in spacetimes with enough symmetries. It is clear that both invariants ${}^{i}\!T^{(T,Q)}_G,{}^{i}\!B^{(T,Q)}_G$ for each set are independently boundary terms in $D=4$. This can be easily seen from Eq.~\eqref{GB-tele-tensorial} and Eq.~\eqref{BGGeneral1} (and Eq.~\eqref{GB-tele-tensorial-alt} with Eq.~\eqref{BGGeneral2}), since the l.h.s. of the split $\lc{G}$ is a boundary term in $D=4$ and both Eq.~\eqref{BGGeneral1} and Eq. \eqref{BGGeneral2} are boundary terms as well. Or in other words, \begin{eqnarray}
^{i}\!T^{(T,Q)}_G=\lc{G}-{}^{i}\!B^{(T,Q)}_G
=\textrm{boundary term}\,.
\end{eqnarray}
It is worth saying that then, this property will hold for any version of TG.

\subsection{Remarks about the known Metric Teleparallel case}

It is worth comparing our general results with the purely Metric TG case originally presented in Ref.~\cite{Kofinas:2014owa}. Since our derivation in Sec.~\ref{sec:derivation} follows theirs, our first set of scalars straightforwardly reduces to the known result when replacing $N_{\mu\nu\rho} \to K_{\mu\nu\rho}$ (vanishing non-metricity) and recalling its anti-symmetry in the first two indices, 
\begin{equation} 
    \lc{G} = {}^1 T^{(T)}_G + {}^{1}\!B^{(T)}_G,
\end{equation}
with
\begin{eqnarray}
    {}^1 T^{(T)}_G &=& \delta^{\mu\,\nu\,\rho\,\sigma}_{\mu_1 \mu_2 \mu_3 \mu_4} \left[ K^{\mu_1}{}_{\alpha\mu}K^{\alpha \mu_2}{}_{\nu}K^{\mu_3}{}_{\beta\rho}K^{\beta \mu_4}{}_{\sigma} + 2 K^{\mu_1 \mu_2}{}_{\mu} K^{\mu_3}{}_{\alpha\nu} K^{\alpha \mu_4}{}_\gamma K^\gamma{}_{\rho\sigma} + 2 K^{\mu_1 \mu_2}{}_{\mu} K^{\mu_3}{}_{\alpha \nu} \nabla_\sigma K^{\alpha \mu_4}{}_{\rho} \right], \\
   {}^{1}\!B^{(T)}_G &= &   \frac{1}{\sqrt{-g}} \, \partial_\mu \! \left[ \sqrt{-g} \, \delta^{\mu \, \nu \, \rho \, \sigma}_{\mu_1 \mu_2 \mu_3 \mu_4} K^{\mu_1 \mu_2}{}_{\nu} \left( K^{\mu_3}{}_{ \lambda \rho} K^{\lambda \mu_4}{}_{\sigma} - \frac{1}{2} \lc{R}^{\mu_3 \mu_4}{}_{\rho \sigma} \right) \right]\, .
\end{eqnarray}
Notice that some terms of the bulk part have been evaluated to zero due to the anti-symmetry, and that the covariant derivative is with respect to the Teleparallel (torsional) connection.

Interestingly, we have also derived an alternative split in Sec.~\ref{sec:derivation-alt} which was not considered in the literature before, but it is as legitimate as any other. When evaluated in the Metric TG case it reads
\begin{equation} 
\lc{G} = {}^{2}T^{(T)}_G+{}^{2}\!B^{(T)}_G\,, 
\end{equation}
with
\begin{eqnarray}
  {}^{2}T^{(T)}_G   &= &\delta^{\mu\,\nu\,\rho\,\sigma}_{\mu_1 \mu_2 \mu_3 \mu_4} \left[ K^{\mu_1}{}_{\alpha\mu}K^{\alpha \mu_2}{}_{\nu}K^{\mu_3}{}_{\beta\rho}K^{\beta \mu_4}{}_{\sigma} - K^{\mu_1}{}_{\alpha\mu}K^{\alpha \mu_2}{}_{\nu} \lc{\nabla}_\sigma K^{\mu_3 \mu_4}{}_{\rho} \right]\,, \\  
 {}^{2}\!B^{(T)}_G&=&-  \frac{1}{2} \frac{1}{\sqrt{-g}} \, \partial_\mu \! \left[ \sqrt{-g} \, \delta^{\mu \, \nu \, \rho \, \sigma}_{\mu_1 \mu_2 \mu_3 \mu_4} K^{\mu_1 \mu_2}{}_{\nu} \lc{R}^{\mu_3 \mu_4}{}_{\rho \sigma} \right]\,.   
\end{eqnarray}
This second novel decomposition may lead to significant differences when compared to known results.

\section{Symmetric Teleparallel Gauss-Bonnet Gravity}\label{sec:STG}
In this section, we focus our study in a particular geometry purely represented by nonmetricity, known as Symmetric TG. Then, we show how the Symmetric TG Gauss-Bonnet invariants behave, and find examples of different spacetimes as well as possible theories to consider.

\subsection{Symmetric Teleparallel Gravity and its Gauss-Bonnet invariant}
Symmetric TG is a particular subset of General TG where torsion is absent, hence gravity being purely represented by nonmetricity. In this case, by taking Eq.~\eqref{RGeneral}, we find that the Levi-Civita scalar can be split as
\begin{eqnarray}
    \lc{R}=Q+B_Q\,,\label{RQB}
\end{eqnarray}
where $Q$ is the so-called nonmetricity scalar defined as
\begin{eqnarray}
    Q=-\,\frac{1}{4}\,Q_{\lambda\mu\nu}Q^{\lambda\mu\nu}+\frac{1}{2}\,Q_{\lambda\mu\nu}Q^{\mu\nu\lambda}+\frac{1}{4}\,Q_{\mu}Q^{\mu}-\frac{1}{2}\,Q_{\mu}\hat{Q}^{\mu}\,, \quad Q_{\mu} \equiv Q_{\mu\nu}\,^{\nu}\,, \quad \hat{Q}_{\mu} \equiv Q_{\nu\mu}\,^{\nu}\,,
\end{eqnarray}
and $B_Q$ a boundary term given by
\begin{eqnarray}
    B_Q=\lc{\nabla}_{\mu}(\hat{Q}^\mu-Q^\mu)\,.
\end{eqnarray}
Then, the General Teleparallel equivalent of GR action written in Eq.~\eqref{GTEGR_action} reduces to the Symmetric Teleparallel equivalent of GR, whose action reads:
\begin{eqnarray}
 	\mathcal{S} _{\rm STEGR} =  \int \dd^4\!x \sqrt{-g} \, Q \,.\label{STEGR}
\end{eqnarray}
One can also consider modified theories of gravity within this type of geometries by changing the above action. In this formalism, since both torsion and curvature are vanishing, the connection can be always written as
\begin{align}\label{CGRGamma}
    \Gamma^{\alpha}{}_{\mu\nu}=\frac{\partial x^{\alpha}}{\partial \xi^{\lambda}}\partial_{\mu}\partial_{\nu}\xi^{\lambda}\,,
\end{align}
where  $\xi^{\alpha}$ is associated to  diffeomorphisms (as a St\"uckelberg field). Then, the connection only has a maximum of 4 d.o.f.. However, it is always possible to find a gauge, known as the coincident gauge, such that the above vector trivializes the connection, i.e. $\Gamma^{\alpha}{}_{\mu\nu}=0$. In that gauge, $\nabla_\mu=\partial_\mu$, and all the d.o.f. of any symmetric teleparallel theory would be encoded in the metric, at the price of loosing diffeomorphism invariance.

Now, let us focus on the Gauss-Bonnet invariant within Symmetric TG. This means replacing $N^\lambda{}_{\mu\nu}\rightarrow L^\lambda{}_{\mu\nu}$ in the expressions obtained in the previous sections. Explicitly, the first set of Symmetric TG Gauss-Bonnet scalar invariants using the expressions~\eqref{TGGeneral1}-\eqref{BGGeneral1} become
\begin{align}
\lc{G} &= {}^{1}T^{(Q)}_G+{}^{1}\!B^{(Q)}_G\,,
\label{GB-tele-tensorial1Q}
\end{align}
where 
\begin{eqnarray}
{}^{1}T^{(Q)}_G&=& \delta^{\mu\,\nu\,\rho\,\sigma}_{\mu_1 \mu_2 \mu_3 \mu_4} \left[ L^{\mu_1}{}_{\alpha\mu}L^{\alpha \mu_2}{}_{\nu}L^{\mu_3}{}_{\beta\rho}L^{\beta \mu_4}{}_{\sigma} -2 L^{\mu_1 \mu_2}{}_{\mu}L^{\mu_3}{}_{\alpha\nu}L^{\alpha}{}_{\beta\rho}L^{[\beta \mu_4]}{}_{\sigma} \right. \notag  \\
     && \qquad \qquad \qquad 
     + 2 g_{\alpha \beta} L^{\mu_1 \mu_2}{}_{\mu} L^{[\mu_3 \alpha]}{}_{\nu} \nabla_\sigma L^{\beta \mu_4}{}_{\rho}  \left. + 4 g_{\alpha \beta} g_{\gamma \delta} L^{\mu_1 \mu_2}{}_{\mu} L^{[\mu_3 \alpha]}{}_\nu L^{\mu_4 \gamma}{}_{\rho} L^{(\delta \beta)}{}_\sigma \right]\,, \label{TG1Sym}\\
{}^{1}\!B^{(Q)}_G&=& \frac{1}{\sqrt{-g}} \, \partial_\mu \! \left[ \sqrt{-g} \, \delta^{\mu \, \nu \, \rho \, \sigma}_{\mu_1 \mu_2 \mu_3 \mu_4} L^{\mu_1 \mu_2}{}_{\nu} \left( L^{\mu_3}{}_{ \lambda \rho} L^{\lambda \mu_4}{}_{\sigma} - \frac{1}{2} \lc{R}^{\mu_3 \mu_4}{}_{\rho \sigma} \right) \right]\,.\label{BG1Sym}
\end{eqnarray}
while using the second split given by~\eqref{TGGeneral2}-\eqref{BGGeneral2} we find
\begin{align}
\lc{G} &={}^{2}T^{(Q)}_G+{}^{2}\!B^{(Q)}_G    \,,   
\label{GB-tele-tensorial2Q}
\end{align}
\begin{eqnarray}
   {}^{2}T^{(Q)}_G&=& \delta^{\mu\,\nu\,\rho\,\sigma}_{\mu_1 \mu_2 \mu_3 \mu_4} \left[ L^{\mu_1}{}_{\alpha\mu}L^{\alpha \mu_2}{}_{\nu}L^{\mu_3}{}_{\beta\rho}L^{\beta \mu_4}{}_{\sigma} - L^{\mu_1}{}_{\alpha\mu}L^{\alpha \mu_2}{}_{\nu} \lc{\nabla}_\sigma L^{\mu_3 \mu_4}{}_{\rho} \right]\,,\label{TG2Sym} \\  
{}^{2}\!B^{(Q)}_G&=&- \frac{1}{2\sqrt{-g}} \, \partial_\mu \! \left[ \sqrt{-g} \, \delta^{\mu \, \nu \, \rho \, \sigma}_{\mu_1 \mu_2 \mu_3 \mu_4} L^{\mu_1 \mu_2}{}_{\nu} \lc{R}^{\mu_3 \mu_4}{}_{\rho \sigma} \right]=- \frac{1}{2} \, \lc{\nabla}_\mu \! \left[ \delta^{\mu \, \nu \, \rho \, \sigma}_{\mu_1 \mu_2 \mu_3 \mu_4} L^{\mu_1 \mu_2}{}_{\nu} \lc{R}^{\mu_3 \mu_4}{}_{\rho \sigma} \right]\,.   \label{BG2Sym}
\end{eqnarray}
Note that the covariant derivative appearing in Eq.~\eqref{TG1Sym} is computed with the Symmetric TG connection. Let us also stress here that the first term on the second line of Eq.~\eqref{TGGeneral1} is identically zero for the Symmetric Teleparallel case, which is why now ${}^{1}T^{(Q)}_G$ only contains four terms.  In the next sections, we will study these scalars in more detail.

\subsection{Spacetimes examples for the Symmetric Teleparallel Gauss-Bonnet invariant}
In this section, we briefly discuss the evaluation of the Symmetric TG Gauss-Bonnet invariants in FLRW cosmology and spherical symmetry.
\subsubsection{FLRW cosmology}\label{sec:FRW}
The line-element for FLRW can be written in spherical coordinates as
\begin{eqnarray}
   \dd s^2=-N(t)^2\dd t^2+\frac{a(t)^2\dd r^2}{1-kr^2}+r^2(\dd \theta^2+\sin^2\theta \, \dd \phi^2)\,,
\end{eqnarray}
where $k$ is the spatial curvature and $N(t)$ and $a(t)$ are the lapse function and scale factor, respectively.
In Ref.~\cite{Hohmann:2021ast}, it was found that depending on how one solves the curvatureless and torsionless conditions, one can get three different branches in flat FLRW cosmology having different connections. However, for the non-flat case $k\neq0$, there is a unique branch satisfying those conditions. The first set of Symmetric TG Gauss-Bonnet invariants given by~\eqref{TG1Sym}-\eqref{BG1Sym} for that case become
\begin{eqnarray}
    {}^{1}T_G^{(Q)}&=&\displaystyle \frac{6 (K-2 H)^2 \dot{H}}{N}+\frac{12 H (K-H) \dot{K}}{N}+6 H^2 \left(-6 H K+4 H^2+3 K^2\right)+\frac{k^2}{a^4}\Big[\frac{6 K \dot{H}+12 (K-H) \dot{K}}{K^3 N}\nonumber\\
    &&-\frac{6 H (H-2 K)}{K^2}\Big]+\frac{k}{a^2}\Big[\frac{\left(\frac{24 H}{K}-12\right) \dot{H}+12 \left(1-\frac{H^2}{K^2}\right) \dot{K}}{N}+\frac{12 H \left(-H K+H^2+K^2\right)}{K}\Big]\,,\\
     {}^{1}\!B_G^{(Q)}&=& \displaystyle \frac{6 K (4 H-K) \dot{H}}{N}+\frac{12 H (H-K) \dot{K}}{N}+18 H^2 K (2 H-K)-\frac{6 k^2 }{a^4 K^2 N}\left(\dot{H}+2 \dot{K}\right)+\frac{12 k^2 H \dot{K}}{a^4 K^3 N}+\frac{6 k^2 H (H-2 K)}{a^4 K^2}\nonumber\\
    &&+\frac{12 k }{a^2}\left(\frac{3 \dot{H}}{N}-H K+3 H^2-\frac{\dot{K}}{N}\right)-\frac{12 k }{a^2 K}\left(\frac{2 H \dot{H}}{N}+H^3\right)+\frac{12 k H^2 \dot{K}}{a^2 K^2 N}\,,
\end{eqnarray}
where dots are differentiation with respect to $t$, $H=\dot{a}/(aN)$ is the Hubble parameter and $K$ represents an extra d.o.f. coming from the connection. The above quantities give us the expected value for the Riemannian Gauss-Bonnet invariant in FLRW cosmology:
\begin{eqnarray}
    {}^{1}T_G^{(Q)}+{}^{1}\!B_G^{(Q)}=\lc{G}=24 H^2 \left(\frac{\dot{H}}{N}+H^2 \right)+\frac{24 k }{a^2}\left(\frac{\dot{H}}{N}+H^2\right)\,.\label{Gform}
\end{eqnarray}
Similarly, the second set of Symmetric TG Gauss-Bonnet invariants expressed by~\eqref{TG2Sym}-\eqref{BG2Sym} for the non-flat FLRW case gives:
\begin{eqnarray}
    {}^{2}T_G^{(Q)}&=&\frac{12 H (K-H) \dot{H}}{N}+\frac{6 H^2 \dot{K}}{N}+6 H^3 (3 K-2 H)+\frac{6k }{a^2}\Big[\frac{2\left(1-\frac{ H}{K}\right) \dot{H}+\left(\frac{ H^2}{K^2}+1\right) \dot{K}}{N}+ H \left(-\frac{H^2}{K}+2 H+K\right)\Big]\nonumber\\
    &&+\frac{6k^2 }{a^4K}\Big[H+\frac{ \dot{K}}{K N}\Big]\,,\\
     {}^{2}\!B_G^{(Q)}&=&\frac{12 H (3 H-K) \dot{H}}{N}-\frac{6 H^2 \dot{K}}{N}+18 H^3 (2 H-K)+\frac{6k }{a^2}\Big[\frac{\frac{2 (H+K) \dot{H}}{K}-\left(\frac{ H^2}{K^2}+1\right) \dot{K}}{N}+ H \left(\frac{H^2}{K}+2 H-K\right)\Big]\nonumber\\
     &&-\frac{6k^2 }{a^4K}\Big[ H+\frac{ \dot{K}}{K N}\Big] \,,
\end{eqnarray}
which also correctly reproduces the Riemannian Gauss-Bonnet invariant~\eqref{Gform}. Since the scalars are different for each set, in principle, one could have different dynamics for a given modified Symmetric TG theory.

Let us now explore the situation in flat FLRW where we have three different branches. Following the notation for the branches introduced in Ref.~\cite{Hohmann:2021ast}, we obtain that the first set of Symmetric TG Gauss-Bonnet scalars become
\begin{eqnarray}
  {}^{1}T_G^{(Q)}&=& \left\{\begin{array}{lr}
    \displaystyle 24 H^2 \left(\frac{\dot{H}}{N}+H^2 \right)\,, & \textrm{First branch}\\
     \displaystyle \frac{6 (2 H+K)^2 \dot{H}}{N}+\frac{12 H (H+K) \dot{K}}{N}+6 H^2 \left(6 H K+4 H^2+3 K^2\right)  \,, & \textrm{Second branch}\\
   \displaystyle \frac{6 (K-2 H)^2 \dot{H}}{N}+\frac{12 H (K-H) \dot{K}}{N}+6 H^2 \left(-6 H K+4 H^2+3 K^2\right) \,, & \textrm{Third branch}
        \end{array}\right.\,\\
         {}^{1}\!B_G^{(Q)}&=& \left\{\begin{array}{lr}
   0\,, & \textrm{First branch}\\
     \displaystyle -\frac{6 K (4 H+K) \dot{H}}{N}-\frac{12 H (H+K) \dot{K}}{N}-18 H^2 K (2 H+K)   \,, & \textrm{Second branch}\\
   \displaystyle \frac{6 K (4 H-K) \dot{H}}{N}+\frac{12 H (H-K) \dot{K}}{N}+18 H^2 K (2 H-K) \,, & \textrm{Third branch}
        \end{array}\right.\,
\end{eqnarray}
and for the second set of Symmetric TG Gauss-Bonnet scalars, we obtain
\begin{eqnarray}
  {}^{2}T_G^{(Q)}&=& \left\{\begin{array}{lr}
    \displaystyle -12 H^2 \left(\frac{\dot{H}}{N}+H^2 \right)\,, & \textrm{First branch}\\
     \displaystyle -\frac{12 H (H+K) \dot{H}}{N}-\frac{6 H^2 \dot{K}}{N}-6 H^3 (2 H+3 K) \,, & \textrm{Second branch}\\
   \displaystyle \frac{12 H (K-H) \dot{H}}{N}+\frac{6 H^2 \dot{K}}{N}+6 H^3 (3 K-2 H) \,, & \textrm{Third branch}
        \end{array}\right.\,\\
         {}^{2}\!B_G^{(Q)}&=& \left\{\begin{array}{lr}
   \displaystyle 36 H^2 \left(\frac{\dot{H}}{N}+H^2 \right)\,, & \textrm{First branch}\\
     \displaystyle \frac{12 H (3 H+K) \dot{H}}{N}+\frac{6 H^2 \dot{K}}{N}+18 H^3 (2 H+K)   \,, & \textrm{Second branch}\\
   \displaystyle \frac{6 H (6 H-2 K) \dot{H}}{N}-\frac{6 H^2 \dot{K}}{N}+18 H^3 (2 H-K) \,. & \textrm{Third branch}
        \end{array}\right.\,
\end{eqnarray}
 Clearly,  for all the branches the combination ${}^{i}T_G^{(Q)}+{}^{i}B_G^{(Q)}$ gives the same value as~\eqref{Gform} with $k=0$, and, as expected, $K$ which is related to the new d.o.f. related to the connection drops out. It is interesting to notice that the scalars for the first set, first branch coincide with the Metric TG case~\cite{Bahamonde:2016kba}. Furthermore, it is also known that in flat FLRW, the first branch, the scalar $Q=6H^2$ also coincides with the torsional scalar case $T=6H^2$. That indicates that the first set, first branch has similar features in both teleparallel versions. Clearly, for the second set, also the first branch has a similar feature since $ {}^{2}B_G^{(Q)}=-3 \,{}^{2}T_G^{(Q)}$, and then, effectively, the dynamics of any theory constructed from those terms would be equivalent to their torsional counterpart at the background level. Let us emphasise here again that our constructed scalars will not have any dynamics in cosmology (or any spacetime) if they appear linearly in an action since they are boundary terms.

\subsubsection{Spherical Symmetry}
As another important spacetime example for displaying the Symmetric TG Gauss-Bonnet scalars, we will choose the following spherically symmetric metric
\begin{eqnarray}
    \dd s^2= -A(r)\dd t^2+\frac{1}{B(r)}\dd r^2+M(r)^2(\dd \theta^2+\sin^2\theta\dd \phi^2)\,,
\end{eqnarray}
where $A,B$ and $M$ are arbitrary functions depending on the radial coordinate. In Ref.~\cite{DAmbrosio:2021zpm}, it was found that there are two branches for spherical symmetry satisfying the torsionless and flat conditions. In Ref.~\cite{Bahamonde:2022esv}, it was found that in the second branch, there is a particular case in which there are exact scalarized solutions for a particular Symmetric TG theory. Let us use that particular situation as an example to show the Symmetric TG Gauss-Bonnet scalars within spherical symmetry. Choosing the same convention as that reference, we find that the first set of Symmetric TG Gauss-Bonnet invariants for the $c=k=0$ case is
\begin{eqnarray}
    {}^{1}T_G^{(Q)}&=&-\frac{2 B\Gamma^r{}_{\theta\theta} A'^2 M'}{A^2M^3}-\frac{2 B^2 A'^2 M'}{A^2M \Gamma^r{}_{\theta\theta}}-\frac{BA'^2 \left(2 BM'^2+1\right)}{A^2M^2}-\frac{B^2 A'^2}{2A^2 (\Gamma^r{}_{\theta\theta})^2}-\frac{(\Gamma^r{}_{\theta\theta})^2 A'^2}{2 A^2M^4}\nonumber\\
    &&+\frac{\Gamma^t{}_{\theta\theta} \left(A'\left(4 B\Gamma^t{}_{\theta\theta} M'-M \left(\Gamma^t{}_{\theta\theta} B'+4 B\Gamma'{}^t{}_{\theta\theta}\right)\right)-2 BM \Gamma^t{}_{\theta\theta} A''\right)}{2 M^5}-\frac{2 B^2 A'\Gamma'{}^r{}_{\theta\theta}}{A(\Gamma^r{}_{\theta\theta})^3}\nonumber\\
    &&+\frac{B}{2 A(\Gamma^r{}_{\theta\theta})^2}\left(2 BA''+\frac{A'\left(3 M B'+4 BM'\left(1-2 \Gamma'{}^r{}_{\theta\theta}\right)\right)}{M}\right)-\frac{(\Gamma^r{}_{\theta\theta})^2 \left(-2 M A''+\frac{M A'B'}{B}+4 A'M'\right)}{2A M^5}\nonumber\\
    &&+\frac{2 M^2 A'B'+B\left(4 M \left(M A''+A'M'\left(3 M B'M'+2 \Gamma'{}^r{}_{\theta\theta}\right)\right)-(\Gamma^t{}_{\theta\theta})^2 A'^2\right)+8 B^2 M^2 M'\left(A''M'+2 A'M''\right)}{2 AM^4}\nonumber\\
    &&+\frac{2 \Gamma^r{}_{\theta\theta} \left(A'\left(\Gamma'{}^r{}_{\theta\theta}-2 BM'^2\right)+M \left(2 BA''M'+A'\left(B'M'+2 BM''\right)\right)\right)}{AM^4}\nonumber\\
    &&+\frac{2 B\left(2 BM A'M''+M'\left(2 BM A''+A'\left(3 M B'+2 BM'\right)\right)\right)}{AM^2 \Gamma^r{}_{\theta\theta}}\,,\\
      {}^{1}\!B_G^{(Q)}&=&\frac{2 B \Gamma^r{}_{\theta\theta} A'^2 M'}{A^2M^3}+\frac{2 B^2 A'^2 M'}{MA^2 \Gamma^r{}_{\theta\theta}}+\frac{3 B A'^2}{A^2M^2}+\frac{B^2 A'^2}{2 A^2(\Gamma^r{}_{\theta\theta})^2}+\frac{(\Gamma^r{}_{\theta\theta})^2 A'^2}{2A^2 M^4}\nonumber\\
      &&+\frac{\Gamma^t{}_{\theta\theta} \left(2 B M \Gamma^t{}_{\theta\theta} A''+A' \left(M \left(\Gamma^t{}_{\theta\theta} B'+4 B \Gamma'{}^t{}_{\theta\theta}\right)-4 B \Gamma^t{}_{\theta\theta} M'\right)\right)}{2 M^5}+\frac{(\Gamma^r{}_{\theta\theta})^2 \left(A' \left(M B'+4 B M'\right)-2 B M A''\right)}{2 B AM^5}\nonumber\\
      &&+\frac{A' \left(B \left((\Gamma^t{}_{\theta\theta})^2 A'-8 M M' \Gamma'{}^r{}_{\theta\theta}\right)-6 M^2 B'\right)}{2A M^4}+\frac{B }{2 A(\Gamma^r{}_{\theta\theta})^2}\left(\frac{A' \left(4 B M' \left(2 \Gamma'{}^r{}_{\theta\theta}-1\right)-3 M B'\right)}{M}-2 B A''\right)\nonumber\\
      &&-\frac{2 \Gamma^r{}_{\theta\theta} \left(A' \left(\Gamma'{}^r{}_{\theta\theta}-2 B M'^2\right)+M \left(2 B A'' M'+A' \left(B' M'+2 B M''\right)\right)\right)}{AM^4}\nonumber\\
      &&-\frac{6 B A''}{AM^2}+\frac{2 B^2 A' \Gamma'{}^r{}_{\theta\theta}}{A(\Gamma^r{}_{\theta\theta})^3}+\frac{-6 B M A' B' M'-4 B^2 \left(M A'' M'+A' \left(M M''+M'^2\right)\right)}{AM^2 \Gamma^r{}_{\theta\theta}}\,,
\end{eqnarray}
where primes are differentiation with respect to $r$. We can easily see that, as expected
\begin{eqnarray}
  {}^{1}T_G^{(Q)}+ {}^{1}\!B_G^{(Q)}=\lc{G}=\frac{4 B A'' \left(B M'^2-1\right)+A' \left(B' \left(6 B M'^2-2\right)+8 B^2 M' M''\right)}{A M^2}-\frac{2 B A'^2 \left(B M'^2-1\right)}{A^2 M^2}   \,.
\end{eqnarray}
Similar to the cosmological case, the connection components only disappear (in a non-trivial way) when one takes the combination that creates the Riemannian  Gauss-Bonnet invariant. One can also compute similar expressions for the second set of scalars, finding that they are also different than the above expressions. We will refrain from writing down their form for simplicity. As a possible application of using those expressions, one can use the minisuperspace approach to find out the spherically symmetric field equations of a particular Symmetric Teleparallel theory of gravity. This can be done by introducing the vector $\xi_\mu$ to replace the connection components and then by computing the Euler-Lagrange equations with respect to $\{A,B,M\}$ to obtain the metric equations and $\xi_\mu$, for the connection equations.

\subsection{New theories constructed with the Symmetric Teleparallel Gauss-Bonnet invariants}
One of the simplest modifications of the Symmetric Teleparallel equivalent of GR that has been studied in the last years relies on promoting $Q$ from appearing linearly as in Eq.~\eqref{STEGR} to an arbitrary function thereof, explicitly
\begin{eqnarray}
 	\mathcal{S} _{f(Q)} =  \int \dd^4\!x \sqrt{-g} \, f(Q)\,.
\end{eqnarray}
This theory has some similar features to its analogous torsional teleparallel version called $f(T)$ gravity~\cite{Ferraro:2006jd,Ferraro:2008ey,Bahamonde:2021gfp} since it has second-order field equations of motions which are dynamically quite different from the Riemannian extension $f(\lc{R})$ which is a fourth-order theory in the metric. Moreover, $f(Q)$ theory has been analysed in cosmology finding that it is compatible with certain cosmological observations~\cite{DAmbrosio:2021pnd,Anagnostopoulos:2021ydo,Frusciante:2021sio}. However, it has been found that the first cosmological branch may potentially suffer from strong-coupling problems~\cite{BeltranJimenez:2019tme}. Further, since $Q$ is related to $\lc{R}$ as~\eqref{RQB}, other more general theories such as $f(Q,B_Q)$ have been considered in order to include $f(\lc{R})$ gravity as a subcase of it~\cite{Gakis:2019rdd,Capozziello:2023vne,Loo:2023uod}.

One can then generalise the above theory by considering the Symmetric TG Gauss-Bonnet invariant derived in the previous section. Obviously, a linear combination of either $T^{(Q)}_G$ or $B^{(Q)}_G$ would not change the dynamics of the above theory since they are boundary terms in 4-dimensions\footnote{For simplicity in the notation, in this section we removed the numbers in the scalars since the discussion does not change for any set of scalars considered.}. Therefore, if one is interested in constructing modified theories of gravity with them, one could for example consider a non-linear action as with the four mentioned scalars:
\begin{eqnarray}
 	\mathcal{S} _{f(Q,B_Q,T^{(Q)}_G,B^{(Q)}_G)} =  \int \dd^4\!x \sqrt{-g} \,f(Q,B_Q,T^{(Q)}_G,B^{(Q)}_G) \,,
\end{eqnarray}
which as a subcase contains the Riemannian modified Gauss-Bonnet theory by setting $f(Q,B_Q,T^{(Q)}_G,B^{(Q)}_G)=f(Q+B_Q,T^{(Q)}_G+B^{(Q)}_G)=f(\lc{R},\lc{G})$ that has been widely studied (see for example~\cite{Nojiri:2005vv,Nojiri:2005jg,Nojiri:2017ncd,Cognola:2006eg}). The theory proposed above, which - in general - is a fourth-order theory in the metric, is also similar to the torsional $f(T,B,B_G^{(T)},T_G^{(T)})$ introduced in Refs.~\cite{Kofinas:2014owa,Bahamonde:2016kba}. As mentioned in the previous section, the first branch of cosmology in flat FLRW has the same scalars as its torsional counterpart. That means that at least at the background level, the theory $f(T,B,T_G^{(T)},B_G^{(T)})$ is equivalent to the first branch of $f(Q,B_Q,T^{(Q)}_G,B^{(Q)}_G) $ for flat FLRW cosmology (they have the same flat FLRW cosmological equations). The reason for that is the fact that for the first branch of flat FLRW, $f(T,B,T_G^{(T)},B_G^{(T)})=f(T,B,T_G^{(T)})=f(Q,B_Q,T^{(Q)}_G,B^{(Q)}_G)=f(Q,B_Q,T^{(Q)}_G)$ (see the scalars and discussion in Sec.~\ref{sec:FRW}). However, this equivalence might be broken at the level of cosmological perturbations. Actually, $f(T,B)$ only has 3 d.o.f. around flat FLRW~\cite{Bahamonde:2020lsm} while $f(Q,B_Q)$ has at least 4 d.o.f. for the first branch~\cite{BeltranJimenez:2019tme}. 

Another route to construct a modified Symmetric TG with a non-trivial contribution from the Symmetric TG Gauss-Bonnet invariants is by introducing a scalar field and couple it non-minimally as
\begin{align}
		\mathcal{S}_{\rm STsGB} 
		&=\frac{1}{2\kappa^2}\int \dd^4\!x \sqrt{-g} \Big[Q-\frac{1}{2}\beta\, \partial_\mu \psi \partial^\mu \psi
		+\alpha_1 \mathcal{G}_1(\psi)T_G^{(Q)}+\alpha_2 \mathcal{G}_2(\psi)B_G^{(Q)}\Big]  \,.\label{scalarSTG}
\end{align}
The above-proposed theory leads to second-order field equations for all the fields (see Appendix~\ref{app:second-order-scalar-STGB}), and contains the so-called scalar Gauss-Bonnet gravity theory in the Riemannian sector by taking the limit $\alpha_1 \mathcal{G}_1(\psi)=\alpha_2 \mathcal{G}_2(\psi)$ which would introduce a coupling of the form $f(\psi)\lc{G}$. The latter has been widely studied in the context of black hole physics since it predicts the existence of scalarized black holes with a spontaneous scalarization process~\cite{Kanti:1995vq,Torii:1996yi,Pani:2009wy,Sotiriou:2014pfa,Silva:2017uqg,Doneva:2017bvd}. Further, two recent studies found the existence of similar solutions in the torsional teleparallel case~\cite{Bahamonde:2022chq,Bahamonde:2023llu} and then, the above Symmetric TG theory could be also potentially interesting to study within that direction. Note that the theory of Eq.~\eqref{scalarSTG} is part of the so-called Symmetric TG Horndeski theory proposed in Ref.~\cite{Bahamonde:2022cmz}, only when one takes the second set ${}^{2}T_G^{(Q)},{}^{2}\!B_G^{(Q)}$ given by Eqs.~\eqref{TG2Sym}-\eqref{BG2Sym}. Although, as mentioned above, both sets satisfy the requirement of second-order field equations, in that study the authors also made the simplifying assumption that there are only terms with up to quadratic contractions of nonmetricity, which includes $\mathcal{G}_2(\psi) \, {}^{2}\!B_G^{(Q)}$ as part of the theory\footnote{Doing one integration by parts, it is easy to see that the resulting operator falls into the classification of Ref.~\cite{Bahamonde:2022cmz} as $N_Q=1$, $n=r=1$ and $m=l=0$, with arbitrary $N_\phi$.}. Furthermore, since the couplings, Eq.~\eqref{scalarSTG} can be always recast as $\alpha_1 \mathcal{G}_1(\varphi)\lc{G}+\alpha \mathcal{G}(\varphi) \, {}^{2}\!B_G^{(Q)}$ and the Symmetric TG Horndeski theory contains the standard Levi-Civita Horndeski part (which contains $\alpha_1 \mathcal{G}_1(\varphi)\lc{G}$), then, the above theory is part of it with the second set. On the other hand, since the two scalars in the first set, ${}^{1}T_G^{(Q)},{}^{1}\!B_G^{(Q)}$ given by Eqs.~\eqref{TG1Sym}-\eqref{BG1Sym}, always contain cubic or quartic contractions of nonmetricity, they are \emph{not} included in the Symmetric TG Horndeski construction of Ref.~\cite{Bahamonde:2022cmz}.

It is, of course, trivial to extend such Symmetric TG theories to the General TG case just by replacing the Symmetric TG Gauss-Bonnet invariants by the general ones from Sec.~\ref{sec:GB}, i.e., $T_G^{(Q)},B_G^{(Q)}\rightarrow T_G^{(T,Q)},B_G^{(T,Q)}$. Those theories will contain both the Metric Gauss-Bonnet theories introduced in Refs.~\cite{Kofinas:2014owa,Bahamonde:2016kba,Bahamonde:2022chq} and also the above proposed Symmetric TG ones.

\section{Summary of results}\label{sec:conclusions}
In this paper, we have found how the Riemannian Gauss-Bonnet invariant is related to torsion and nonmetricity in the General TG framework. We found that it is possible to define newly constructed Teleparallel Gauss-Bonnet invariants that are always boundary terms in $D=4$ dimensions from which the Riemannian Gauss-Bonnet emerges. Our result coincides with the result found in Ref.~\cite{Kofinas:2014owa} when nonmetricity is vanishing, which is related to torsional Gauss-Bonnet gravity. We then focus on the new terms that are only related to nonmetricity and express all our quantities in Symmetric TG. 

We found that the way of splitting the Riemannian Gauss-Bonnet invariant in terms of teleparallel quantities is not unique. We present two different ways of splitting it with different bulk and boundary terms. After this, we focused on the Symmetric TG case and we analysed how these scalars behave in non-flat FLRW cosmology and also in spherical symmetry. Since those terms are boundary terms in 4-dimensions, if we want them to acquire dynamics, one would need to consider them in an action non-linearly. We did this by formulating two new Symmetric TG theories constructed with the Gauss-Bonnet invariants where we allowed more general modifications of the STEGR action~\eqref{STEGR}. One of those theories contains as a special case, the so-called modified Gauss-Bonnet gravity theory constructed in the Riemannian sector and where the action is constructed as $f(\lc{R},\lc{G})$. The second proposed theory contains the so-called scalar-Gauss-Bonnet theory where the Gauss-Bonnet acquires dynamics by coupling it with a scalar field.

In future studies, it would be interesting to analyse further the proposed theories with the Symmetric TG invariants. They might serve as a starting point to study cosmology in Symmetric TG and also to explore new routes in the process of understanding the difference between the torsional and nonmetricity versions of Teleparallel gravity. Furthermore, scalarized black holes with spontaneous scalarization process should exist in~\eqref{scalarSTG}, but potentially, they might have different features as their Riemannian or torsional counterparts due to the fact that now, there are more possible ways of solving the teleparallel condition in spherical symmetry which introduces richer dynamics. Furthermore, the connection components are still dynamical in this formalism, and one could have different black hole configurations while having different Symmetric Teleparallel connections. It would be interesting to analyse how the connection affects the dynamics of such black holes and also their thermodynamics.

Another route that might lead to interesting applications is to follow a similar approach as~\cite{Glavan:2019inb} and construct a non-trivial $D=4$ linear Teleparallel Gauss-Bonnet gravity theory. This can be done by following the construction of~\cite{Lu:2020iav,Hennigar:2020lsl} which is to take the Teleparallel GB invariants and use a dimensional reduction method.

Recently, an interesting and related work appeared \cite{Bajardi:2023gkd}. Although the authors study the Gauss-Bonnet invariant in General TG theories of gravity, which is also this work's primary goal, here we focus on and exploit the power of differential form language to obtain a compact expression for our scalars. Moreover, our formalism relies on always rewriting the invariants as boundary terms (or topological invariants\footnote{Note that we are not showing that our scalars are topological invariants separately, but they are always boundary terms in 4-dimensions.}), while in Ref.~\cite{Bajardi:2023gkd}, the authors explore a different split of the Riemannian Gauss-Bonnet invariant into pieces that are not independently boundary terms. For example, Eq.~(14) in that reference provides an expression for the torsional case in terms of three scalars. Considering the Lagrangian $L=\sqrt{-g}(c_1\mathcal{G}_T+c_2\mathcal{G}_{\mathcal{D}\mathcal{T}}+c_3\mathcal{G}_{\mathcal{D}\mathcal{D}\mathcal{T}})$, the only way of having boundary terms without dynamics is taking the combination $c_3=c_2=c_1$, which reconstructs the Riemannian Gauss-Bonnet invariant. A similar situation occurs for the nonmetricity Gauss-Bonnet invariant (given by Eq.~(21) in~\cite{Bajardi:2023gkd}). Thus, none of the scalars are boundary terms (or topological invariants) on their own, but only the unique combination reconstructing the Riemannian Gauss-Bonnet term is. In that sense, our approach also differs from the one of those authors' since our split always keeps a structure such that the Teleparallel scalars are divided into a bulk and a boundary term; and in 4-dimensions, our Teleparallel Gauss-Bonnet terms always correspond to boundary terms independently when they appear linearly in an action.

\section*{Acknowledgements}
The work of JMA has been supported by CONICET, ANPCyT and UBA. JMA would like to thank the support and funding of UBA to carry out the current work by means of the grant ``Financiamiento de estad\'ia en el exterior 2023", and to the Institute of Physics of the Czech Academy of Sciences (CEICO) for funding and hospitality during initial stages of this work.
S.B. was supported by JSPS Postdoctoral Fellowships for Research in Japan and KAKENHI Grant-in-Aid for Scientific Research No. JP21F21789, and by ``Agencia Nacional de Investigaci\'on y Desarrollo" (ANID), Grant ``Becas Chile postdoctorado al extranjero" No. 74220006.
The work of G.T. was supported by the Grant Agency of the Czech Republic, GACR grant
20-28525S. The work of L.G.T. was supported by European Union (Grant No. 101063210). S.B., G.T., and L.G.T. would like to thank the support by the Bilateral Czech-Japanese Mobility
Plus Project JSPS-21-12 (JPJSBP120212502).
S.B. was also supported by “Agencia Nacional de Investigación y Desarrollo” (ANID), Grant “Becas Chile postdoctorado al extranjero” No. 74220006.

\appendix 

\section{Useful properties of differential forms}

This is an incomplete list of properties that are used in the main text and Appendix \ref{app:equiv} to work on the expressions in the differential form language. First, consider the Levi-Civita convariant exterior derivative $\lc{\DD}$, then it follows 
\begin{eqnarray} 
    \lc{\DD} e^a &=& 0, \label{lc-D-e} \\
    \lc{\DD} \eta_{ab} &=& 0. \label{lc-D-eta}
\end{eqnarray}
We also have the Riemannian differential Bianchi identity
\begin{eqnarray} \label{lc-Bianchi-R}
    \lc{\DD} \lc{R}^{a}{}_{b} &=& 0. 
\end{eqnarray}
In an orthonormal frame we also have the property
\begin{equation} \label{lc-D-epsilon}
    \lc{\DD} \epsilon_{a_1...a_D} = \dd \epsilon_{a_1...a_D} = 0,
\end{equation}
where $\dd$ is the usual exterior derivative. Notice that with Eq.~\eqref{lc-D-eta} we can raise the index in Eq.~\eqref{curvature-split} 
\begin{equation}\label{curvature-split2}
    R^{ab} = \lc{R}^{ab} + \lc{D} N^{ab} + N^a{}_c \wedge N^{cb}.
\end{equation}
Acting twice with $\lc{\DD}$ gives the usual relation with the Riemannian \textit{curvature 2-form}
\begin{equation} \label{D2-N-up-down}
\lc{\DD}^2 N^a{}_b = \lc{R}^a{}_c \wedge N^c{}_b - \lc{R}^c{}_b \wedge N^a{}_c.
\end{equation}
Consider now the exterior covariant derivative for the teleparallel connection $\DD$,
\begin{eqnarray}
\DD \eta_{ab} &=& - 2 \omega_{(ab)} = - 2 N_{(ab)},\\
\DD \eta^{ab} &=& 2 N^{(ab)}.
\end{eqnarray}
Raising an index under $\DD$ is now nontrivial when the \textit{distortion 1-form} has a symmetric part, i.e. nonmetricity,
\begin{eqnarray} \label{D-N-up-up-to-up-down}
\DD N^{ab} &=& \eta^{bc} \DD N^{a}{}_{c} - 2 N^{a}{}_c \wedge N^{(cb)}.
\end{eqnarray}
We can trade $\DD$ for the Levi-Civita one $\lc{\DD}$ plus an extra term
\begin{equation}
    \DD N^a{}_b = d N^a{}_b + \omega^a{}_c \wedge N^c{}_b - \omega^c{}_b \wedge N^a{}_c = \lc{D} N^a{}_b + 2 N^a{}_c \wedge N^c{}_b,
\end{equation}
or, equivalently
\begin{equation} \label{D-N-up-up}
    \DD N^{ab} = \lc{\DD} N^{ab} + 2 N^a{}_c \wedge N^{[cb]}.
\end{equation}
This allows us to reexpress Eq.~\eqref{curvature-split} in terms of $\DD$,
\begin{eqnarray} \label{curvature-split3}
    R^{a}{}_{b} = \lc{R}^{a}{}_{b} + \DD N^{a}{}_{b} - N^{a}{}_c \wedge N^{c}{}_{b}.
\end{eqnarray}

\section{Equivalence of alternative decompositions} \label{app:equiv}

The purpose of this appendix is to prove the equivalence of Eqs.~\eqref{GB-tele} and \eqref{GB-tele-alt}. We work our way starting from Eq.~\eqref{GB-tele-alt} and integrating by parts the first term of the second line, effectively altering the split between bulk and boundary,
\begin{eqnarray}
    \epsilon_{a_1...a_D} N^{a_1}{}_f \wedge N^{f a_2} \wedge \lc{\DD} N^{a_3 a_4} \wedge e^{a_5} \wedge \dots \wedge e^{a_D} &=& \epsilon_{a_1...a_D} \lc{\DD} N^{a_1 a_2} \wedge N^{a_3}{}_f \wedge N^{f a_4} \wedge e^{a_5} \wedge \dots \wedge e^{a_D} \notag \\    
    &=& \epsilon_{a_1...a_D}  \dd \left( N^{a_1 a_2} \wedge N^{a_3}{}_f \wedge N^{f a_4} \wedge e^{a_5} \wedge \dots \wedge e^{a_D} \right) \notag \\
    &&+ \epsilon_{a_1...a_D} N^{a_1 a_2} \wedge \lc{\DD}( N^{a_3}{}_f \wedge N^{f a_4}) \wedge e^{a_5} \wedge \dots \wedge e^{a_D},
\end{eqnarray}    
which, when plugged back in leads to
\begin{eqnarray} \label{GB-tele4}
    \lc{G} \, *\!1 &=& \frac{1}{(D-4)!} \epsilon_{a_1...a_D} \Biggl[ - \dd \left( 2 N^{a_1 a_2} \wedge \lc{R}^{a_3 a_4} \wedge e^{a_5} \wedge \dots \wedge e^{a_D} + N^{a_1 a_2} \wedge \lc{\DD} N^{a_3 a_4} \wedge e^{a_5} \wedge \dots \wedge e^{a_D} \right) \notag \\    
    && \qquad\qquad\qquad\qquad + N^{a_1 a_2} \wedge \lc{\DD}( N^{a_3}{}_f \wedge N^{f a_4}) \wedge e^{a_5} \wedge \dots \wedge e^{a_D} \notag \\
    && \qquad\qquad\qquad\qquad + N^{a_1}{}_f \wedge N^{f a_2} \wedge N^{a_3}{}_h \wedge N^{h a_4} \wedge e^{a_5} \wedge \dots \wedge e^{a_D} \Biggr],
\end{eqnarray}
where we also used $N^{a_3}{}_f \wedge N^{f a_4} = - \lc{R}^{a_3 a_4} - \lc{\DD} N^{a_3 a_4}$. The boundary term here already looks the same as in Eq.~\eqref{GB-tele}, but we still need to massage the bulk part, in particular the second line above, 
\begin{eqnarray}
&& \epsilon_{a_1...a_D} N^{a_1 a_2} \wedge \lc{\DD}( N^{a_3}{}_f \wedge N^{f a_4}) \wedge e^{a_5} \wedge \dots \wedge e^{a_D} \notag \\
&=& \epsilon_{a_1...a_D} \eta_{cd} N^{a_1 a_2} \wedge \left( \lc{\DD}N^{a_3 c} \wedge N^{d a_4} - N^{a_3 c} \wedge \lc{\DD} N^{d a_4} \right) \wedge e^{a_5} \wedge \dots \wedge e^{a_D} \notag \\
&=& \epsilon_{a_1...a_D} \eta_{cd} N^{a_1 a_2} \wedge \left( \lc{\DD}N^{a_3 c} \wedge N^{d a_4} + \lc{\DD} N^{c a_3} \wedge N^{a_4 d} \right) \wedge e^{a_5} \wedge \dots \wedge e^{a_D} \notag \\
&=& \epsilon_{a_1...a_D} \eta_{cd} N^{a_1 a_2} \wedge 2 \left( \lc{\DD} N^{(a_3 c)} \wedge N^{(d a_4)} + \lc{\DD} N^{[a_3 c]} \wedge N^{[d a_4]} \right) \wedge e^{a_5} \wedge \dots \wedge e^{a_D}\,,
\end{eqnarray}
where in the last equality we used $N^{a b} = N^{(a b)} + N^{[a b]}$. From property \eqref{D-N-up-up}, we have
\begin{eqnarray}
    \lc{\DD} N^{(a_3 c)} &=& \cancelto{0}{\DD N^{(a_3 c)}} - \left( N^{a_3}{}_f \wedge N^{[fc]} + N^c{}_f \wedge N^{[f a_3]} \right), \\
    \lc{\DD} N^{[a_3 c]} &=& \DD N^{[a_3 c]} - \left( N^{a_3}{}_f \wedge N^{[fc]} - N^c{}_f \wedge N^{[f a_3]} \right),
\end{eqnarray}
where we used the third Bianchi identity of Eq.~\eqref{eq:Bianchi} in the teleparallel case, which reads $\DD N^{(ab)} = 0$ (recalling $N^{(ab)} = Q^{ab}$). Then, we have
\begin{eqnarray}
\lc{\DD} N^{(a_3 c)} \wedge N^{(d a_4)} + \lc{\DD} N^{[a_3 c]} \wedge N^{[d a_4]} &=& - \eta_{fh} N^{a_3 f} \wedge N^{[hc]} \wedge N^{(d a_4)} -  \eta_{fh} N^{c f} \wedge N^{[h a_3]} \wedge N^{(d a_4)} \\
&& + \DD N^{[a_3 c]} \wedge N^{[d a_4]} - \eta_{fh} N^{a_3 f} \wedge N^{[h c]} \wedge N^{[d a_4]} + \eta_{fh} N^{c f} \wedge N^{[h a_3]} \wedge N^{[d a_4]} \,, \notag
\end{eqnarray}
and therefore, replacing back and reordering we get
\begin{eqnarray}
\epsilon_{a_1...a_D} N^{a_1 a_2} \wedge \lc{\DD}( N^{a_3}{}_f \wedge N^{f a_4}) \wedge e^{a_5} \wedge \dots \wedge e^{a_D} &=& 2 \epsilon_{a_1...a_D} \eta_{cd} N^{a_1 a_2} \wedge \DD N^{[a_3 c]} \wedge N^{[d a_4]} \wedge e^{a_5} \wedge \dots \wedge e^{a_D} \\
&-& \notag 2 \epsilon_{a_1...a_D} \eta_{cd} \eta_{fh} N^{a_1 a_2} \wedge N^{a_3 f} \wedge N^{[hc]} \wedge N^{(d a_4)} \wedge e^{a_5} \wedge \dots \wedge e^{a_D} \notag \\ 
&-& \notag 2 \epsilon_{a_1...a_D} \eta_{cd} \eta_{fh} N^{a_1 a_2} \wedge N^{a_3 f} \wedge N^{[hc]} \wedge N^{[d a_4]} \wedge e^{a_5} \wedge \dots \wedge e^{a_D} \notag \\ 
&-& \notag 2 \epsilon_{a_1...a_D} \eta_{cd} \eta_{fh} N^{a_1 a_2} \wedge N^{c f} \wedge N^{[h a_3]} \wedge N^{(d a_4)} \wedge e^{a_5} 
\wedge \dots \wedge e^{a_D} \notag \\ 
&+& \notag 2 \epsilon_{a_1...a_D} \eta_{cd} \eta_{fh} N^{a_1 a_2} \wedge N^{c f} \wedge N^{[h a_3]} \wedge N^{[d a_4]} \wedge e^{a_5} \wedge \dots \wedge e^{a_D}. \notag
\end{eqnarray}
Now here notice that the last 4 terms (those without $\DD$) contain three factors of $N$ contracted to each other directly, of the form
\begin{equation}
    \epsilon_{\dots a_3 a_4 \dots} \eta_{cd} \eta_{fh} N^{a_3 c} \wedge N^{d f} \wedge N^{h a_4}.
\end{equation}
Due to the presence of the totally antisymmetric tensor $\epsilon_{\dots a_3 a_4 \dots}$, only the antisymmetric part w.r.t. $a_3$ and $a_4$ of the above triple $N$ survives. By splitting each $N$ into symmetric and antisymmetric parts we can see which combinations survive. In particular, we find
\begin{eqnarray}\label{eq:vanishing-combin}
    \epsilon_{a_1...a_D} \eta_{cd} \eta_{fh} N^{(a_3 f)} \wedge N^{[hc]} \wedge N^{(d a_4)} &=& 0, \\ 
    \epsilon_{a_1...a_D} \eta_{cd} \eta_{fh} N^{[a_3 f]} \wedge N^{[hc]} \wedge N^{[d a_4]} &=& 0, \label{eq:vanishing-combin-2} \\ 
    \epsilon_{a_1...a_D} \eta_{cd} \eta_{fh} N^{[c f]} \wedge N^{[h a_3]} \wedge N^{[d a_4]} &=& 0, 
\end{eqnarray}
which leads to
\begin{eqnarray}
\epsilon_{a_1...a_D} N^{a_1 a_2} \wedge \lc{\DD}( N^{a_3}{}_f \wedge N^{f a_4}) \wedge e^{a_5} \wedge \dots \wedge e^{a_D} &=& 2 \epsilon_{a_1...a_D} \eta_{cd} N^{a_1 a_2} \wedge \DD N^{[a_3 c]} \wedge N^{[d a_4]} \wedge e^{a_5} \wedge \dots \wedge e^{a_D} \\
&-& \notag 2 \epsilon_{a_1...a_D} \eta_{cd} \eta_{fh} N^{a_1 a_2} \wedge N^{[a_3 f]} \wedge N^{[hc]} \wedge N^{(d a_4)} \wedge e^{a_5} \wedge \dots \wedge e^{a_D} \notag \\ 
&-& \notag 2 \epsilon_{a_1...a_D} \eta_{cd} \eta_{fh} N^{a_1 a_2} \wedge N^{(a_3 f)} \wedge N^{[hc]} \wedge N^{[d a_4]} \wedge e^{a_5} \wedge \dots \wedge e^{a_D} \notag \\ 
&-& \notag 2 \epsilon_{a_1...a_D} \eta_{cd} \eta_{fh} N^{a_1 a_2} \wedge N^{(c f)} \wedge N^{[h a_3]} \wedge N^{(d a_4)} \wedge e^{a_5} 
\wedge \dots \wedge e^{a_D} \notag \\ 
&-& \notag 2 \epsilon_{a_1...a_D} \eta_{cd} \eta_{fh} N^{a_1 a_2} \wedge N^{[c f]} \wedge N^{[h a_3]} \wedge N^{(d a_4)} \wedge e^{a_5} 
\wedge \dots \wedge e^{a_D} \notag \\ 
&+& \notag 2 \epsilon_{a_1...a_D} \eta_{cd} \eta_{fh} N^{a_1 a_2} \wedge N^{(c f)} \wedge N^{[h a_3]} \wedge N^{[d a_4]} \wedge e^{a_5} \wedge \dots \wedge e^{a_D}. \notag
\end{eqnarray}
Here it is immediate to see  that the second and fifth lines cancel each other out. Then, after some reordering and relabeling we get
\begin{eqnarray}
\epsilon_{a_1...a_D} N^{a_1 a_2} \wedge \lc{\DD}( N^{a_3}{}_f \wedge N^{f a_4}) \wedge e^{a_5} \wedge \dots \wedge e^{a_D} &=& 2 \epsilon_{a_1...a_D} \eta_{fh} N^{a_1 a_2} \wedge \DD N^{[a_3 f]} \wedge N^{[h a_4]} \wedge e^{a_5} \wedge \dots \wedge e^{a_D} \\
&-& \notag 2 \epsilon_{a_1...a_D} \eta_{cd} \eta_{fh} N^{a_1 a_2} \wedge N^{(a_3 f)} \wedge N^{[hc]} \wedge N^{[d a_4]} \wedge e^{a_5} \wedge \dots \wedge e^{a_D} \notag \\ 
&+& \notag 2 \epsilon_{a_1...a_D} \eta_{cd} \eta_{fh} N^{a_1 a_2} \wedge N^{[a_3 f]} \wedge N^{(hc)} \wedge N^{[d a_4]} \wedge e^{a_5} 
\wedge \dots \wedge e^{a_D} \notag \\ 
&+& \notag 2 \epsilon_{a_1...a_D} \eta_{cd} \eta_{fh} N^{a_1 a_2} \wedge N^{(a_3 f)} \wedge N^{(hc)} \wedge N^{[d a_4]} \wedge e^{a_5} \wedge \dots \wedge e^{a_D}. \notag
\end{eqnarray}
Using Eq.~\eqref{D-N-up-up-to-up-down} it simplifies to,
\begin{eqnarray}
\epsilon_{a_1...a_D} N^{a_1 a_2} \wedge \lc{\DD}( N^{a_3}{}_f \wedge N^{f a_4}) \wedge e^{a_5} \wedge \dots \wedge e^{a_D} &=& 2 \epsilon_{a_1...a_D} N^{a_1 a_2} \wedge \DD N^{a_3}{}_{c} \wedge N^{[c a_4]} \wedge e^{a_5} \wedge \dots \wedge e^{a_D}  \\
&-& 2 \epsilon_{a_1...a_D} \eta_{cd} \eta_{fh} N^{a_1 a_2} \wedge N^{a_3 f} \wedge N^{(h c)} \wedge N^{[d a_4]} \wedge e^{a_5} \wedge \dots \wedge e^{a_D} \notag \\
&-& \notag 2 \epsilon_{a_1...a_D} \eta_{cd} \eta_{fh} N^{a_1 a_2} \wedge N^{(a_3 f)} \wedge N^{[hc]} \wedge N^{[d a_4]} \wedge e^{a_5} \wedge \dots \wedge e^{a_D}\,,\notag
\end{eqnarray}
while using Eq.~\eqref{eq:vanishing-combin-2} allows us to combine the last two terms
\begin{eqnarray}
&& \epsilon_{a_1...a_D} \eta_{cd} \eta_{fh} \left( N^{a_3 f} \wedge N^{(h c)} \wedge N^{[d a_4]} + N^{(a_3 f)} \wedge N^{[h c]} \wedge N^{[d a_4]} \right) \notag \\
&=& \epsilon_{a_1...a_D} \eta_{cd} \eta_{fh} \left( N^{a_3 f} \wedge N^{(h c)} \wedge N^{[d a_4]} + N^{a_3 f} \wedge N^{[h c]} \wedge N^{[d a_4]} \right) \notag \\
&=& \epsilon_{a_1...a_D} \eta_{cd} \eta_{fh} \left( N^{a_3 f} \wedge N^{h c} \wedge N^{[d a_4]} \right).
\end{eqnarray}
Finally, we arrived at the following identity
\begin{eqnarray} \label{some-diff-form-identity}
\epsilon_{a_1...a_D} N^{a_1 a_2} \wedge \lc{D}( N^{a_3}{}_f \wedge N^{f a_4}) \wedge e^{a_5} \wedge \dots \wedge e^{a_D} &=& 2 \epsilon_{a_1...a_D} N^{a_1 a_2} \wedge D N^{a_3}{}_{c} \wedge N^{[c a_4]} \wedge e^{a_5} \wedge \dots \wedge e^{a_D} \\
&-& 2 \epsilon_{a_1...a_D} N^{a_1 a_2} \wedge N^{a_3}{}_{d} \wedge N^{d}{}_{c} \wedge N^{[c a_4]} \wedge e^{a_5} \wedge \dots \wedge e^{a_D}, \notag
\end{eqnarray}
which plugged back into Eq.~\eqref{GB-tele4} leads to precisely to Eq.~\eqref{GB-tele}.

\section{Second order nature of scalar-ST-Gauss-Bonnet} \label{app:second-order-scalar-STGB}

Here we want to prove that the field equations associated to the theory Eq.~\eqref{scalarSTG} are second order in derivatives of all fields $(\psi, g_{\mu\nu}, \Gamma^{\alpha}{}_{\mu\nu})$, for the two splits of $\lc{G}$ proposed in Sec.~\ref{sec:GB} and evaluated in the Symmetric TG case in Eqs.~\eqref{GB-tele-tensorial1Q} and \eqref{GB-tele-tensorial2Q}. In the following it is useful to recall that, from its definition in Eqs.~\eqref{disformation-tensor} and \eqref{nonmetricity-tensor}, the disformation tensor contains up to first derivatives of the metric and no derivatives of the Teleparallel connection, schematically
\begin{equation} \label{disformation-tensor-derivs}
    L \sim \partial g - \Gamma.
\end{equation}

As explained in the main text, we can always recast the last two terms in Eq.~\eqref{scalarSTG} as $\alpha_1 \mathcal{G}_1(\psi)\lc{G}+\alpha \mathcal{G}(\psi) \, B_G^{(Q)}$, where we have absorbed $T_G^{(Q)}$ into $\lc{G}$ and redefined the coupling of $B_G^{(Q)}$. Since $\alpha_1 \mathcal{G}_1(\psi)\lc{G}$ belongs to Riemannian Horndeski \cite{Kobayashi:2011nu}, it is already guaranteed to lead to second-order field equations, and therefore it suffices to prove that it is also true for $\alpha \mathcal{G}(\psi) \, B_G^{(Q)}$ for each split. Moreover, the difference between the two boundary terms associated with each split trivially leads to second-order field equations
\begin{eqnarray}
    \int d^4\!x \sqrt{-g} \mathcal{G}(\psi) \left[ {}^{2}\!B^{(Q)}_G - {}^{1}\!B^{(Q)}_G \right] &=& \int d^4\!x \sqrt{-g} \mathcal{G}'(\psi) \partial_\mu \psi \, \delta^{\mu \, \nu \, \rho \, \sigma}_{\mu_1 \mu_2 \mu_3 \mu_4} L^{\mu_1 \mu_2}{}_{\nu} L^{\mu_3}{}_{ \lambda \rho} L^{\lambda \mu_4}{}_{\sigma} \,,
\end{eqnarray}
where we integrated by parts once, as it contains at most first derivatives of any given field. Therefore, it only remains to be shown that any of the two possible boundary terms leads to second-order field equations. We focus on $\mathcal{G}(\psi) \, {}^{2}\!B^{(Q)}_G$. 

\subsection{Scalar field equation}

First notice that the bulk terms ${}^i T_G^{(Q)}$ of both splits, Eqs.~\eqref{TG1Sym} and \eqref{TG2Sym} respectively, contain at most second derivatives of the fields. Then, given that the Riemannian Gauss-Bonnet invariant $\lc{G}$ itself also has this property, it follows that the corresponding boundary terms ${}^i\! B_G^{(Q)}$, given by Eqs.~\eqref{BG1Sym} and \eqref{BG2Sym}, also have at most second derivatives. Therefore, variations w.r.t. the scalar filed $\psi$ are automatically second order in this theory for both splits, and in particular, for $\mathcal{G}(\psi) \, {}^{2}\!B^{(Q)}_G$.

\subsection{Teleparallel connection equation}

We continue by integrating $\mathcal{G}(\psi) \, {}^{2}\!B^{(Q)}_G$ by parts
\begin{equation} \label{scalar-2B_G}
    \int d^4\!x \sqrt{-g} \mathcal{G}(\psi)\,{}^2\!B_G^{(Q)} = \frac{1}{2} \int d^4\!x \sqrt{-g} \mathcal{G}'(\psi) \partial_\mu \psi \, \delta^{\mu \, \nu \, \rho \, \sigma}_{\mu_1 \mu_2 \mu_3 \mu_4} L^{\mu_1 \mu_2}{}_{\nu} \lc{R}^{\mu_3 \mu_4}{}_{\rho \sigma},
\end{equation}
which is a second order expression. Given that the Teleparallel connection $\Gamma^{\alpha}{}_{\mu\nu}$ appears here only linearly and without any derivatives acting on it, as per Eq.~\eqref{disformation-tensor-derivs}, it follows then that variations w.r.t. to it are automatically second order too.

\subsection{Metric field equations}

Finally, let us consider variations w.r.t. the metric, which are the less trivial ones. Indeed, by counting derivatives in Eq.~\eqref{scalar-2B_G}, we get schematically $\partial \psi \, \partial g \, \partial^2 g$, which could potentially lead to higher-than-second order field equations. Let us work on the variation explicitly in order to show that those higher-order terms actually cancel,
\begin{equation}
    \delta_g \int d^4\!x \sqrt{-g} \mathcal{G}(\psi)\,{}^2\!B_G^{(Q)} \supset \frac{1}{2} \int d^4\!x \sqrt{-g} \, \partial_\mu \psi \, \delta^{\mu \, \nu \, \rho \, \sigma}_{\mu_1 \mu_2 \mu_3 \mu_4} \delta_g \left[ L^{\mu_1 \mu_2}{}_{\nu} \lc{R}^{\mu_3 \mu_4}{}_{\rho \sigma} \right] \equiv \mathcal{E}_g,
\end{equation}
where we are not tracking terms which lead automatically to second-order field equations. It is useful to recall that under the Teleparallel condition, Eq.~\eqref{eq:tel-cond}, from Eq.~\eqref{BB9} it is possible to write
\begin{equation}
    \lc{R}^{\sigma}\,_{\rho\mu\nu} = \lc{\nabla}_{\nu}L^{\sigma}\,_{\mu\rho} - \lc{\nabla}_{\mu}L^{\sigma}\,_{\nu\rho} + \dots = 2 \lc{\nabla}_{[\nu}L^{\sigma}\,_{\mu]\rho} + \dots,
\end{equation}
where we also dropped terms with less than two derivatives acting on the metric, which are safe. The above implies
\begin{equation}
    \delta_g \lc{R}^{\sigma}\,_{\rho\mu\nu} = 2 \lc{\nabla}_{[\nu} \delta L^{\sigma}\,_{\mu]\rho}, 
\end{equation}
and therefore
\begin{equation}
    \mathcal{E}_g = \int d^4\!x \sqrt{-g} \, \partial_\mu \psi \, \delta^{\mu \, \nu \, \rho \, \sigma}_{\mu_1 \mu_2 \mu_3 \mu_4} \left[ \delta_g L^{\mu_1 \mu_2}{}_{\nu} \lc{\nabla}_{\sigma} L^{\mu_3\mu_4}\,_{\rho} + L^{\mu_1 \mu_2}{}_{\nu} \lc{\nabla}_{\sigma} \delta_g L^{\mu_3\mu_4}\,_{\rho} \right],
\end{equation}
where we used the antisymmetry of $\rho$ and $\sigma$ to drop the antisymmetrization brackets, and we also used the symmetry of $L$ in the last two indices. Let us work on the second term to show that contributes the same as the first. We first integrate by parts the second term
\begin{eqnarray}
    \mathcal{E}_g &=& \int d^4\!x \sqrt{-g} \, \partial_\mu \psi \, \delta^{\mu \, \nu \, \rho \, \sigma}_{\mu_1 \mu_2 \mu_3 \mu_4} \left[ \delta_g L^{\mu_1 \mu_2}{}_{\nu} \lc{\nabla}_{\sigma} L^{\mu_3\mu_4}\,_{\rho} - \delta_g L^{\mu_3\mu_4}\,_{\rho} \lc{\nabla}_{\sigma} L^{\mu_1 \mu_2}{}_{\nu} \right] \notag \\
    &-& \int d^4\!x \sqrt{-g} \, \lc{\nabla}_{\sigma} \lc{\nabla}_\mu \psi \, \delta^{\mu \, \nu \, \rho \, \sigma}_{\mu_1 \mu_2 \mu_3 \mu_4} L^{\mu_1 \mu_2}{}_{\nu} \delta_g L^{\mu_3\mu_4}\,_{\rho}. 
\end{eqnarray}
Notice that the second term with two derivatives acting on $\psi$ is zero due to antisymmetry of $\mu$ and $\sigma$. We then do an exchange and relabel of $\rho$ and $\nu$ (one minus sign) and a cyclic permutation of $\mu_1 \mu_2 \mu_3 \mu_4$ (no sign change),
\begin{eqnarray}
    \mathcal{E}_g &=& \int d^4\!x \sqrt{-g} \, \delta^{\mu \, \nu \, \rho \, \sigma}_{\mu_1 \mu_2 \mu_3 \mu_4} \left[ \delta_g L^{\mu_1 \mu_2}{}_{\nu} \lc{\nabla}_{\sigma} L^{\mu_3\mu_4}\,_{\rho} + \delta_g L^{\mu_1\mu_2}\,_{\nu} \lc{\nabla}_{\sigma} L^{\mu_3 \mu_4}{}_{\rho} \right] \notag \\
    &=& 2 \int d^4\!x \sqrt{-g} \, \delta^{\mu \, \nu \, \rho \, \sigma}_{\mu_1 \mu_2 \mu_3 \mu_4} \delta_g L^{\mu_1 \mu_2}{}_{\nu} \lc{\nabla}_{\sigma} L^{\mu_3\mu_4}\,_{\rho} \notag \\
    &=& 2 \int d^4\!x \sqrt{-g} \, \delta^{\mu \, \nu \, \rho \, \sigma}_{\mu_1 \mu_2 \mu_3 \mu_4} \delta_g L^{\mu_1 \mu_2}{}_{\nu} \lc{R}^{\mu_3 \mu_4}{}_{\rho \sigma},
\end{eqnarray}
where in the last step we again have the Levi-Civita Riemann tensor, up to terms that do not have more that second deriviatives of the metric. 

We finally now express $\delta_g L^{\mu_1 \mu_2}{}_{\nu}$ in terms of $\delta g_{\mu\nu}$, 
\begin{eqnarray}
    \delta_g L^{[\mu_1 \mu_2]}{}_{\nu} = 2 g^{\alpha [\mu_1} g^{\mu_2] \beta} \lc{\nabla}_\alpha \delta g_{\beta\nu},
\end{eqnarray}
which allows us to pull out another derivative and actually see the dangerous term:
\begin{eqnarray}
    \mathcal{E}_g &=& 4 \int d^4\!x \sqrt{-g} \, \delta^{\mu \, \nu \, \rho \, \sigma}_{\mu_1 \mu_2 \mu_3 \mu_4}  \lc{R}^{\mu_3 \mu_4}{}_{\rho \sigma} g^{\alpha \mu_1} g^{\mu_2 \beta} \lc{\nabla}_\alpha \delta g_{\beta\nu} = - 4 \int d^4\!x \sqrt{-g} \, \delta^{\mu \, \nu \, \rho \, \sigma}_{\mu_1 \mu_2 \mu_3 \mu_4} g^{\mu_2 \beta} \delta g_{\beta\nu} \lc{\nabla}^{\mu_1} \lc{R}^{\mu_3 \mu_4}{}_{\rho \sigma}.
\end{eqnarray}
The last factor contains the potentially dangerous third derivatives of $g$. However, because of total antisymmetrization of $\mu_1\mu_3\mu_4$, it is proportional to
\begin{eqnarray}
\lc{\nabla}^{[\mu_1} \lc{R}^{\mu_3 \mu_4]}{}_{\rho \sigma} = 0,
\end{eqnarray}
which vanishes due to the second Bianchi identity, proving the absence of higher-than-second order terms in the field equations.

\bigskip
\bigskip
\noindent

\bibliographystyle{utphys}
\bibliography{references}

\end{document}